\ifdefined\TITClassLoaded
\else
\ifdefined\TITReview
\documentclass[journal,onecolumn]{IEEEtran}
\else
\documentclass[journal]{IEEEtran}
\fi
\fi
\usepackage[cmex10]{amsmath}
\usepackage{amssymb,amsthm}
\usepackage{array,booktabs}
\usepackage{url}
\ifdefined\TITReview
\newenvironment{roadmaptable}{\begin{table}[!ht]}{\end{table}}
\newenvironment{proofmapfigure}{\begin{figure}[!ht]}{\end{figure}}
\else
\newenvironment{roadmaptable}{\begin{table*}[t]}{\end{table*}}
\newenvironment{proofmapfigure}{\begin{figure*}[t]}{\end{figure*}}
\fi
\newtheorem{theorem}{Theorem}
\newtheorem{proposition}[theorem]{Proposition}
\newtheorem{lemma}[theorem]{Lemma}
\newtheorem{corollary}[theorem]{Corollary}
\theoremstyle{remark}
\newtheorem{remark}[theorem]{Remark}
\newtheorem{example}[theorem]{Example}
\newcommand{\F}{\mathbb F}
\newcommand{\Sym}{\operatorname{Sym}}
\newcommand{\coker}{\operatorname{coker}}
\newcommand{\rank}{\operatorname{rank}}
\newcommand{\supp}{\operatorname{supp}}
\newcommand{\Span}{\operatorname{span}}
\newcommand{\ord}{\operatorname{ord}}
\newcommand{\eps}{\varepsilon}
\newcommand{\A}{\mathcal A}

\allowdisplaybreaks[1]

\title{MSR Codes With Polynomial Subpacketization\\
via Symmetric Algebra and Multivariate Interpolation}
\author{Ke Wang\textsuperscript{\textdagger}, Wenjie Ma\textsuperscript{\textdaggerdbl,*},
Zhipeng Li\textsuperscript{\textsection}, Yaqiang Zhang\textsuperscript{\textdagger},
Xiao Li\textsuperscript{\textdagger}, and Guiwen Luo\textsuperscript{\textdagger}%
\thanks{\raggedright\textsuperscript{\textdagger}Ke Wang, Yaqiang Zhang, Xiao Li, and Guiwen Luo are with IEIT SYSTEMS Co., Ltd.,
Jinan 250101, China
(e-mail: \texttt{\{wangke12, zhangyaqiang, lixiao09, luoguiwen\}@ieisystem.com}).}%
\thanks{\raggedright\textsuperscript{\textdaggerdbl}Wenjie Ma is with the School of Mathematics,
Southwest Minzu University, Chengdu 610225, China
(e-mail: \texttt{mawenjie@swun.edu.cn}).}%
\thanks{\raggedright\textsuperscript{\textsection}Zhipeng Li is with the School of Computing and Information Science,
Fuzhou Institute of Technology, Fuzhou 350506, China
(e-mail: \texttt{lzp13484@gmiot.com}).}%
\thanks{\raggedright\textsuperscript{*}Corresponding author: Wenjie Ma.}}
\date{}
\begin{document}
\maketitle
\begin{abstract}
Duursma and Wang proposed a $t$-fold symmetric-algebra framework
extending product-matrix codes to $(n,k,d)$ minimum storage regenerating
(MSR) codes (where $d$ is the number of helper nodes and
$r=d-k+1\geq2$) with parameters $d=tr$, $k=(t-1)r+1$,
length-independent subpacketization $\alpha=\binom{(t-1)r}{t-1}$, and
per-helper download $\alpha/r$, but checked existence only numerically
over the integers for $\alpha\leq10011$, leaving an algebraic proof
open. Casting exact repair as a multivariate symmetric interpolation
problem, we characterize exact repair by the nonvanishing of a square
repair determinant and prove that for $t=3,4$ and every integer
$r\geq2$, this determinant is a nonzero polynomial in every
characteristic. For any $n\geq d+1$, this
yields all-node exact-repair MSR codes with $\alpha=r(2r-1)$ ($t=3$)
and $\alpha=\binom{3r}{3}$ ($t=4$), rates approaching $2/3$ and $3/4$ at
fixed $n-d$, and shortened codes for all
$d\geq\lceil t(k-1)/(t-1)\rceil$ ($k<d<n$). At $n=d+1$ in fixed
characteristic, deterministic polynomial-time search constructs valid
codes over $O(n^2)$ and $O(n^4)$ fields, while helper-set interpolation
gives closed-form binary codes over extension fields of polynomial
degree (order $2^{\mathrm{poly}(n)}$). Compared with existing explicit
constructions, $\alpha$ drops substantially: e.g., from $64$ to $28$ at
$(n,k,d)=(13,9,12)$ ($t=3,r=4$) and from $38416$ to $11480$ at
$(57,43,56)$ ($t=4,r=14$), versus $256$ and $537824$ with optimal
access. Finally, at $r=2$ ($d=k+1$), we establish all-characteristic
realization for two further constant-subpacketization families: $t=5$
($k=9$, $d=10$, any $n\geq11$, $\alpha=70$) and $t=6$ ($k=11$, $d=12$,
any $n\geq13$, $\alpha=252$), achieving rates up to $9/11$ and $11/13$.
\end{abstract}

\section{Introduction}\label{sec:intro}
Distributed storage systems employ erasure coding to maintain data
availability across unreliable storage nodes~\cite{survey}. Storing a
file of $k\alpha$ symbols over a finite field $F$ in an $(n,k)$
maximum-distance-separable (MDS) array code places $\alpha$ symbols on
each of $n$ nodes so that any $k$ nodes suffice to reconstruct the file,
achieving storage rate $R=k/n$ and tolerating up to $n-k$ simultaneous
node failures.

Whereas file reconstruction is infrequent, single-node repair is a
routine background operation. Rebuilding a failed node by downloading
full contents from $k$ survivors transfers $k\alpha$ symbols to restore
only $\alpha$ symbols. Regenerating codes reduce this repair bandwidth
by allowing a replacement node to contact $d$ surviving helpers
($k\leq d<n$) and download a compressed $\beta$-symbol linear message
from each~\cite{dimakis}. At minimum storage per node, the cut-set bound
gives $d\beta\geq d\alpha/(d-k+1)$, and a \emph{minimum storage
regenerating} (MSR) code achieves equality $\beta=\alpha/r$, where
$r:=d-k+1$.
Here $n-k=r+(n-1-d)$, which equals $r$ at the full-helper point $d=n-1$.
We study linear \emph{exact-repair} MSR codes, in which every failed
node is reconstructed identically from any $d$-subset of survivors, with
particular emphasis on small subpacketization $\alpha$, moderate finite
field size $q$, and explicit node coordinates (without imposing the
optimal-access constraint that each helper read only $\beta$ stored
symbols).

\subsection{Subpacketization regimes and the determinant problem}
Existing explicit MSR constructions fall into two distinct scaling
regimes with respect to subpacketization, summarized in
Table~\ref{tab:guarantees}.

\begin{roadmaptable}
\centering\small\setlength{\tabcolsep}{3.5pt}
\caption{Comparison of explicit exact-repair MSR code constructions for
$r=d-k+1\geq2$ (with $\beta=\alpha/r$). All field inequalities in the
fifth column are sufficient conditions; $D_3$ and $D_a$ denote the
shortest-length thresholds in \eqref{eq:shortfields}, and
$Q_{\rm LWHY}=\widehat n r+r2^{r-1}$ in \eqref{eq:lwhyfield} (at
$(t,r)=(4,2)$, fixed MDS second factors require only $q>D_x=136$).}
\label{tab:guarantees}
\begin{tabular}{>{\raggedright\arraybackslash}p{0.20\textwidth}>{\raggedright\arraybackslash}p{0.16\textwidth}>{\raggedright\arraybackslash}p{0.12\textwidth}>{\raggedright\arraybackslash}p{0.18\textwidth}>{\raggedright\arraybackslash}p{0.16\textwidth}>{\raggedright\arraybackslash}p{0.07\textwidth}}
\toprule
Construction & Regime $(k,d)$ & $\alpha$ ($n\geq d+1$) & $\alpha$ at $n=tr+1$ & Field $q$ ($n=tr+1$) & Repair \\
\midrule
Rashmi et al.~\cite{pm} ($t=2$) & $d\geq2k-2$ & $r$ & $r$ & $q>r(n-1)$\textsuperscript{a} & All \\
Ye--Barg~\cite{ybaccess} & $d=n-1$\textsuperscript{b} & $r^{\lceil n/r\rceil}$ & $r^{t+1}$ & $q\geq r\lceil n/r\rceil$ & All \\
WTB~\cite{wtb} (shortened) & $d=n-1$ & $r^{\lceil k/(r+1)\rceil}$ & $r^{\lceil((t-1)r+1)/(r+1)\rceil}$ & Suff.\ large~\cite{wtb} & Syst. \\
Wang et al.~\cite{wlhy} & $d=k+1$ ($r=2$) & $2^{\lceil n/3\rceil}$ & $2^{\lceil(2t+1)/3\rceil}$ & $O(n)$~\cite{wlhy} & All \\
LWHY~\cite{lwhy} & $k+1\leq d<n$ & $r^{\lceil n/(r+1)\rceil}$ & $r^3$ ($t=3$); $r^4$ ($t=4,r\geq3$) & $q\geq\widehat n r+r2^{r-1}$ & All \\
LTS~\cite{lts2026} & $k+1\leq d<n$ & $r^{\lceil n/2\rceil}$ & $r^{\lceil(tr+1)/2\rceil}$ & $q>r\lceil n/2\rceil$\textsuperscript{c} & All \\
\midrule
Here, $t=3$ (Thm.~\ref{thm:main}) & $2d\geq3(k-1)$ & $r(2r-1)$ & $r(2r-1)<r^3$ ($\forall r\geq2$) & $q>D_3=O(r^2)$ & All \\
Here, $t=4$ (Thm.~\ref{thm:main}) & $3d\geq4(k-1)$ & $\binom{3r}{3}$ & $\binom{3r}{3}<r^4$ ($\forall r\geq4$) & $q>D_a=O(r^4)$ & All \\
Here, $t=5,6$ (Cor.~\ref{cor:highercode}) & $d=k+1$, $k\leq2t-1$ & $70,\ 252$ & $70$ ($n=11$); $252$ ($n=13$) & $q>560,\ 2304$ & All \\
\bottomrule
\multicolumn{6}{@{}p{0.93\textwidth}@{}}{\footnotesize\textsuperscript{a}Vandermonde realization of~\cite{pm}; see Section~\ref{subsec:t2pm}. \textsuperscript{b}Full-helper regime; smaller $d$ in~\cite{ybaccess} restricts helper groups (see~\cite{smalld} for arbitrary helpers). \textsuperscript{c}Requires $\gcd(r,q-1)=1$. Shortening applies to the last row by applying the shortening argument of Corollary~\ref{cor:shorteneddegree} to the MSR codes of Corollary~\ref{cor:highercode}.}
\end{tabular}
\end{roadmaptable}

In the first regime, when each helper is restricted to optimal access
(reading only $\beta$ symbols), Balaji et~al.~\cite{bvk} proved the
lower bound $\alpha\geq r^{\lceil n/r\rceil-1}$ at $d=n-1$, matched up
to a factor of $r$ at $d=n-1$ by Ye and Barg~\cite{ybaccess}
($\alpha=r^{\lceil n/r\rceil}$, with grouped-helper extensions for
$d<n-1$) and refined for arbitrary $d$-helper sets by Vajha
et~al.~\cite{smalld}; relaxing all-node repair to systematic nodes yields
$\alpha=r^{\lceil k/(r+1)\rceil}$ in Wang, Tamo, and Bruck~\cite{wtb}.
Dropping optimal access permits smaller exponents while retaining
all-node repair: Wang et~al.~\cite{wlhy} obtain
$\alpha=2^{\lceil n/3\rceil}$ at $r=2$ ($d=k+1$); Li et~al.~\cite{lwhy}
achieve $\alpha=r^{\lceil n/(r+1)\rceil}$ across all $k+1\leq d<n$ over
fields of order $q\geq\widehat n r+r2^{r-1}$; and Li, Tang, and
Song~\cite{lts2026} give $\alpha=r^{\lceil n/2\rceil}$ over linear
fields with $\gcd(r,q-1)=1$. When $k/r$ grows without bound, exponential
subpacketization remains unavoidable even without optimal
access~\cite{ag}. Nevertheless, in the general-$d$ constructions of this
first class~\cite{wlhy,lwhy,lts2026}, fixing $(k,d)$ and increasing the
code length $n$ (adding parity nodes at constant repair degree) forces
$\alpha$ to grow exponentially in $n$; moreover, the field threshold
of~\cite{lwhy} contains an $r2^{r-1}$ term that grows exponentially in
$r$.

In the second regime, multilinear and determinant-based architectures
decouple $\alpha$ completely from the code length $n$. Determinant
codes~\cite{elyasimohajer2019} ($d=k$) and cascade
codes~\cite{elyasimohajer2020} (together with their multilinear
formulation~\cite{duursmaliwang2020}) achieve length-independent
exact repair over $O(n)$ fields across general $(k,d)$, with MSR
subpacketization $\alpha=r^k$ exponential in $k$. When $k/r$ is bounded
by an integer $t-1$ ($t\geq2$), symmetric-algebra constructions reduce
$\alpha$ further to a polynomial of degree $t-1$ in $r$. At $t=2$, the
product-matrix MSR construction of Rashmi, Shah,
and Kumar~\cite{pm} realizes parameters $(k,d)=(r+1,2r)$ with
$\alpha=r$ and $\beta=1$ (and all $d\geq2k-2$ by shortening, covering
asymptotic rates up to $1/2$), subsequently cast in bivariate symmetric
interpolation form by Eid and Duursma~\cite[Sec.~3]{eidduursma2016}.
Duursma and Wang~\cite{atrahasis} extended the product-matrix
architecture to $t$-fold symmetric algebra (termed \emph{Atrahasis codes}
in~\cite{atrahasis}), with parameters $d=tr$ and $k=(t-1)r+1$,
second-factor dimension $m=(t-1)(r-1)+1$, and subpacketization and
per-helper download
\begin{equation}\label{eq:alphabeta}
\alpha=\binom{(t-1)r}{t-1},\qquad
\beta=\binom{(t-1)r-1}{t-2},\qquad\alpha=r\beta.
\end{equation}
At $t=2$ these formulas recover $(d,\alpha,\beta)=(2(k-1),r,1)$. Passing
to $t=3,4$ lowers the repair degree to $3(k-1)/2$ and $4(k-1)/3$
(raising storage rates toward $2/3$ and $3/4$ at fixed $n-d$) while
keeping $\alpha=O_t(r^{t-1})$ polynomial in $r$ and independent of $n$.
Patra and Barg~\cite[Sec.~V]{patrabarg2025} formulated generalized
product-matrix repair in symmetric algebra conditional on a
helper-subspace spanning condition; polynomial-coefficient storage,
layered interpolation reconstruction, and a $t=2$ repair example also
appear in our preliminary patent disclosure~\cite{teamPatent2025}
(see~\cite{krickszanto2026} for symmetric Hermite normal forms; below we
introduce characteristic-free Hermite jet coordinates tailored to the
$t=4$ repair matrix).

In this symmetric-algebra framework, for a fixed MDS node configuration,
exact repair from a $d$-helper set $H$ under the prescribed linear
downloads holds if and only if the evaluated $t\alpha\times t\alpha$
repair determinant $h(x_1,y_1,\ldots,x_d,y_d)$ is nonzero at those node
coordinates (Propositions~\ref{prop:code}
and~\ref{prop:repairnecessity}); existence of a valid configuration over
all sufficiently large fields of a given characteristic therefore reduces
to proving that $h$ is nonzero \emph{as a polynomial} over that prime
field~\cite[Sec.~5.1, Prop.~24]{atrahasis}. Duursma and
Wang~\cite[p.~721 and Prop.~23]{atrahasis} verified $h\ne0$ over
$\mathbb Z$ by randomized finite-field evaluation for all instances with
$\alpha\leq10011$ and posed Open Problem~5: \emph{``Prove or disprove
that the determinant is nonzero for all $\alpha$.''} For fixed $t\geq3$,
two issues arise: the proof must handle unbounded $r\geq2$, and it must
hold in \emph{every} characteristic. Establishing nonvanishing in each
characteristic is strictly stronger than nonvanishing over $\mathbb Z$,
since an integer determinant can vanish identically modulo small primes,
and polarization identities acquire factorial denominators in
characteristic two.

\subsection{Main results and proof overview}
To our knowledge, this paper gives the first uniform proof that the
repair determinant is a nonzero polynomial across unbounded $r\geq2$ and
every characteristic for $t=3,4$. This resolves Open Problem~5
of~\cite{atrahasis} for both infinite families and realizes the
length-independent subpacketization \eqref{eq:alphabeta}, accompanied by
explicit finite-field bounds and node-selection algorithms. For $t=3$,
and also at $r=2$ for $t=4,5,6$, we prove a stronger statement than
\cite[Prop.~24]{atrahasis}: for \emph{any} prescribed MDS family of
second-factor vectors $y_h$, the repair determinant is a nonzero
polynomial in the first factors $x_h$ alone. For $t=4$ and general
$r\geq2$, we take $y_h=y(a_h)$ on a Vandermonde curve and prove joint
nonvanishing in $(x_h,a_h)$. In particular, this implies nonvanishing of
the unrestricted polynomial $h(x_1,y_1,\ldots,x_d,y_d)$ while reducing
the second factor of each node to a single scalar. Multiplying all
reconstruction and helper-set determinants yields one finite-field
configuration serving every failed node and every $d$-helper set.

\begin{theorem}[Uniform repair-matrix realization]\label{thm:main}
Let $t\in\{3,4\}$, $r\geq2$, and $n\geq tr+1$. There are linear
all-node exact-repair minimum storage regenerating codes with $d=tr$,
$k=(t-1)r+1$, and \eqref{eq:alphabeta} over every finite field of order
$q>D$, where
\begin{align}\label{eq:mainbounds}
t=3:\quad D&=\binom{n-1}{2}+(2r-1)\binom{n-1}{3r-1},\\
t=4:\quad D&=\max\{D_x,D_a\},\nonumber\\
D_x&=\binom{n-1}{3}+\binom{3r-1}{2}\binom{n-1}{4r-1},\nonumber\\
D_a&=n-1+(3r-3)\binom{3r-1}{2}\binom{n-1}{4r-1}.\nonumber
\end{align}
For $t=3$, any prescribed $n$ second factors with every $m$ independent
can be used; the same holds at $(t,r)=(4,2)$ over every field of order
$q>D_x$. For $t=4$ and general $r\geq2$, the construction selects
distinct Vandermonde parameters together with the first factors. In each
case one configuration serves every failed node and every set of $d$
surviving helpers.
\end{theorem}

\begin{proof}
The determinant assertions are Theorems~\ref{thm:t3} and~\ref{thm:t4}
and Proposition~\ref{prop:t4r2}. Theorems~\ref{thm:field3}
and~\ref{thm:field4} select a common point where every required
determinant is nonzero, and Proposition~\ref{prop:code} gives
reconstruction and exact repair.
\end{proof}

Once the $n$ node coordinates are fixed, any failed node can be repaired
from any of the $\binom{n-1}{d}$ helper sets. While Theorem~\ref{thm:main}
treats $d=tr$ and $k=(t-1)r+1$, standard MSR shortening extends it to a
broad design region.

\begin{corollary}[Repair degrees obtained by shortening]\label{cor:shorteneddegree}
Fix $t\in\{3,4\}$, $1\leq k\leq d<n$, and $r=d-k+1\geq2$.
If $(t-1)d\geq t(k-1)$, there is an all-node exact-repair MSR code
with parameters $(n,k,d)$, subpacketization and per-helper download
given by \eqref{eq:alphabeta}, and file size $k\alpha$.
The field bound \eqref{eq:mainbounds} applies at code length
$n+s$, where $s=(t-1)d-t(k-1)$.
\end{corollary}
\begin{proof}
Apply the standard MSR shortening operation
\cite[Thm.~6 and Cor.~7]{pm}, \cite[Sec.~1.2, Lemmas~6--7]{atrahasis}.
Start from the MSR code of Theorem~\ref{thm:main} with length $n+s$,
dimension parameter $k+s=(t-1)r+1$, and repair degree $d+s=tr$.
Since $(n+s)-(d+s)=n-d\geq1$, Theorem~\ref{thm:main} applies.
Use any $k+s$ nodes as systematic coordinates, constrain $s$
of them to store zero, and delete them. This leaves $k\alpha$ free
file symbols. Any $k$ surviving nodes, together with the $s$ known
zero nodes, reconstruct the file. For repair, adjoin the same
zero nodes to any $d$ available helpers. Their linear downloads are
zero and can be supplied locally. The $d$ contacted helpers each send
$\beta=\alpha/r$, meeting the cut-set bound because $d-k+1=r$.
\end{proof}

For fixed $n,d$ with $2\leq d<n$, Corollary~\ref{cor:shorteneddegree}
covers every integer $1\leq k\leq k_{\max}$ with rate $R=k/n$, where
\begin{equation}\label{eq:degree-rate}
k_{\max}=\min\!\left\{d-1,\,1+\left\lfloor\frac{(t-1)d}{t}\right\rfloor\right\},
\qquad R_{\max}=\frac{k_{\max}}{n}.
\end{equation}
Along $d=tr$, $R=1/n+((t-1)/t)(d/n)$ approaches $2/3$ for
$t=3$ and $3/4$ for $t=4$ at fixed $n-d$. For example, at $n=9$, the
$(t,r)=(4,2)$ MSR code has $(k,d)=(7,8)$ and rate $7/9$, whereas
shortening one node of a $(10,7,8)$ MSR code gives $(k,d)=(6,7)$ and rate
$2/3$ at the same $(\alpha,\beta)=(20,10)$. As shown in
Table~\ref{tab:guarantees}, at shortest length $n=tr+1$ our
subpacketization beats~\cite{lwhy} for every $r\geq2$ at $t=3$
($r(2r-1)<r^3$) and every $r\geq4$ at $t=4$ ($\binom{3r}{3}<r^4$, with
$84$ versus $81$ at $r=3$). Along the same line, the exponential
$r2^{r-1}$ field term of~\cite{lwhy} is replaced by $D_3=O(r^2)$ and
$D_a=O(r^4)$. When $(k,d)$ are fixed and $n$ grows, $\alpha$ stays
constant and the field bound in \eqref{eq:mainbounds} grows as
$O_{t,r}(n^{tr-1})$. Both $\alpha$ and the per-node bit granularity
$\alpha\log_2 q$ therefore improve upon exponential-in-$n$ constructions
once $n-d$ grows, with $\alpha$ already smaller for $n\geq13$ at $(4,2)$
and $n\geq17$ at $(4,3)$. For dense $y_f$, however, each helper reads
all $\alpha$ stored symbols (Section~\ref{sec:comparison}).

To test how far the paired-channel method extends at $r=2$ ($d=k+1$),
where baseline blocks have pure two-form Koszul kernels,
Subsection~\ref{subsec:higherr2} (Theorem~\ref{thm:higherrepair} and
Corollary~\ref{cor:highercode}) and Appendices~\ref{app:t5}--\ref{app:t6}
establish fixed-MDS nonvanishing in every characteristic for $t=5$
($\alpha=70$) and $t=6$ ($\alpha=252$). These two instances yield
constant-$\alpha$ codes of shortest-length rates $9/11$ and $11/13$ over
fields of orders $q>560$ and $q>2304$.

For coordinate construction at $n=d+1$ in fixed characteristic $p$,
Theorem~\ref{thm:detselection} gives a deterministic polynomial-time
search over fields of order $O_p(n^2)$ for $t=3$ and $O_p(n^4)$ for
$t=4$; the search remains polynomial time for any fixed excess $n-d$.
Without candidate search, Theorem~\ref{thm:interpolatednodes} gives
closed-form binary coordinates over $\F_{2^b}$ with $b$ polynomial in
$n$ at fixed $n-d$ (field order $2^{\mathrm{poly}(n)}$). On the negative
side, Theorem~\ref{thm:curveobstruction} and
Corollary~\ref{cor:curveimpossibility} show that consecutive-power
first-factor curves $(1,z_h,\ldots,z_h^{t-1})$ are singular and prevent
exact repair under the prescribed linear downloads for all
$t\geq3,r\geq2$. Constructing closed-form coordinates directly over
polynomial-size fields remains open.

The remainder of the paper is organized as follows.
Section~\ref{sec:model} presents the polynomial-coefficient encoder, the
subset-interpolation decoder, and the $t=2$ product-matrix case.
Section~\ref{sec:structure} establishes the unified perturbation,
Koszul-syzygy, and Schur-lifting framework used across all determinant
proofs. Sections~\ref{sec:t3} and~\ref{sec:t4} prove uniform
nonvanishing for $t=3$ and $t=4$ across all $r\geq2$, together with the
fixed-MDS $r=2$ extensions for $t=4,5,6$ (with the $t=5,6$ proofs given
in Appendices~\ref{app:t5}--\ref{app:t6}). Section~\ref{sec:fields}
gives deterministic small-field selection, Las Vegas sampling, and
closed-form interpolated binary coordinates.
Sections~\ref{sec:obstruction} and~\ref{sec:comparison} prove the
consecutive-power obstruction and compare subpacketization, field size,
access, and decoding cost, and Section~\ref{sec:conclusion} concludes
with open problems for general $t$.

\section{The Repair Matrix and Its Interpolation Meaning}\label{sec:model}
\subsection{File symbols, node storage, and helper transmissions}
For Vandermonde nodes the encoder admits a scalar polynomial description.
Represent the file by $t$ symmetric polynomials
$P_0,\ldots,P_{t-1}$ in $T_1,\ldots,T_t$, each of degree less than $m$
in each variable. A file symbol is the common coefficient of the
monomials obtained by permuting one exponent tuple. Node $h$ stores
the independent coefficients of
\[
C_h(T_2,\ldots,T_t)=\sum_{i=0}^{t-1}x_{h,i}P_i(a_h,T_2,\ldots,T_t).
\]
To repair node $f$, helper $h$ evaluates $T_2=a_f$ and sends the
independent coefficients of the result. The following coordinate
formulas also cover arbitrary second-factor vectors.

All spaces are over a field $F$. Let $X=F^t$ and let $Y$ be the
span of independent symbols $z_0,\ldots,z_{m-1}$. We use the ordinary
monomial space $\Sym^jY$, indexed by
\[
\Omega_j=\{\delta\in\mathbb Z_{\geq0}^m:|\delta|=j\},\qquad
|\Omega_j|=\binom{m+j-1}{j}.
\]
\emph{File and node storage.}
A file is a linear functional
$a$ on $X\otimes\Sym^tY$, specified by
$a_{i,\delta}=a(e_i\otimes z^\delta)$; its size is
$t\binom{m+t-1}{t}=k\alpha$.
For $x_h\in X$ and a linear form $y_h\in Y$, node $h$ stores $a$ on
$V_h=x_h\otimes y_h\Sym^{t-1}Y$. In coordinates this is
\begin{equation}\label{eq:nodecoefficients}
c_{h,\gamma}=\sum_{i=0}^{t-1}\sum_{b=0}^{m-1}
x_{h,i}y_{h,b}a_{i,\gamma+e_b},\qquad\gamma\in\Omega_{t-1}.
\end{equation}
For nonzero $x_h,y_h$, multiplication by $y_h$ is injective, so these
are $\alpha$ independent stored functionals.

\emph{Repair of node $f$.}
Let $H$ be any $d$ surviving nodes. Helper $h\in H$ uses the public
vector $y_f$ and its own stored symbols to send the $\beta$ scalars
\begin{equation}\label{eq:downloadcoefficients}
d_{h,\gamma}=\sum_b y_{f,b}c_{h,\gamma+e_b},
\qquad\gamma\in\Omega_{t-2}.
\end{equation}
These are $a$ on $x_h\otimes y_hy_f\Sym^{t-2}Y$.
In matrix form the encoder and the helper each perform a multiplication.
Let $M=(a_{i,\delta})$ be the $t$-row file matrix. Write $G_j(y)$
for multiplication by $y$ from $\Sym^jY$ to $\Sym^{j+1}Y$ in the
monomial bases: its column indexed by $\gamma$ contains the
coefficients of $y z^\gamma$. The row stored at node $h$ and the
row it sends to repair $f$ are, respectively,
\begin{align*}
\mathbf c_h &= x_h^{\mathsf T} M G_{t-1}(y_h),\\
\mathbf d_{h\to f} &= \mathbf c_h G_{t-2}(y_f).
\end{align*}
All helpers use the same right-hand matrix for a given failure.
The replacement decodes their rows using the transpose of
\begin{equation}\label{eq:repairmap}
\begin{aligned}
\A_H&:\bigoplus_{h\in H}\Sym^{t-2}Y\longrightarrow X\otimes\Sym^{t-1}Y,\\
&(q_h)\longmapsto\sum_{h\in H}x_h\otimes y_hq_h.
\end{aligned}
\end{equation}
It is square of order $d\beta=t\alpha$. For a given failed node $f$,
the downloads determine its contents for every file precisely when
\begin{equation}\label{eq:targetcriterion}
x_f\otimes\Sym^{t-1}Y\ \subseteq\ \operatorname{im}\A_H .
\end{equation}
Indeed, the helpers give the values of the file functional on
$y_f\operatorname{im}\A_H$, and the requested values are its
restriction to $V_f$. A functional is determined on $V_f$ by those
values exactly when $V_f\subseteq y_f\operatorname{im}\A_H$.
Multiplication by the nonzero form $y_f$ is injective, giving
\eqref{eq:targetcriterion}. Invertibility of $\A_H$ ensures this
inclusion simultaneously for all failed nodes outside $H$.
\subsection{Decoding the helper transmissions}
The same file can be written using independent variable groups.
For an order-$j$ array $c$ and an ordered tuple
$\mathbf b=(b_1,\ldots,b_j)$, put
$\delta(\mathbf b)=\sum_\ell e_{b_\ell}$ and define
\begin{equation}\label{eq:polarpoly}
\begin{split}
&B_c(U;Z^{(1)},\ldots,Z^{(j)})\\
&\quad=\sum_{i=0}^{t-1}\sum_{\mathbf b\in\{0,\ldots,m-1\}^j}
U_i c_{i,\delta(\mathbf b)}\prod_{\ell=1}^j Z^{(\ell)}_{b_\ell}.
\end{split}
\end{equation}
This form is linear in each group and symmetric in its $Z$ groups.
Basis-vector substitutions recover every coefficient in every
characteristic: in characteristic two,
$Z^{(1)}_0Z^{(2)}_1+Z^{(1)}_1Z^{(2)}_0$ vanishes on the diagonal
$Z^{(1)}=Z^{(2)}$ yet remains recoverable from independent variable
groups. The coefficient pairing with the ordinary monomial space is
$\langle c,\sum_i e_i\otimes q_i\rangle
=\sum_{i,\delta}c_{i,\delta}[z^\delta]q_i$.

For the file array $a$ of order $t$, node storage becomes the
independent coefficients of
\begin{equation}\label{eq:nodepoly}
C_h(Z^{(2)},\ldots,Z^{(t)})
=B_a(x_h;y_h,Z^{(2)},\ldots,Z^{(t)}).
\end{equation}
Define
$Q_f(U;Z^{(1)},\ldots,Z^{(t-1)})
=B_a(U;Z^{(1)},\ldots,Z^{(t-1)},y_f)$.
The downloads in \eqref{eq:downloadcoefficients} are exactly the
independent coefficients of
\begin{equation}\label{eq:downloadpoly}
D_{h\to f}=Q_f(x_h;y_h,Z^{(2)},\ldots,Z^{(t-1)}).
\end{equation}
For an arbitrary order-$(t-1)$ array $c$, collect these restrictions
as $\mathcal E_Hc$. Its rows are $(h,\gamma)$ and its columns
$(i,\delta)$, with
\begin{equation}\label{eq:interpolationentry}
(\mathcal E_H)_{(h,\gamma),(i,\delta)}
=\sum_{b:\,\delta=\gamma+e_b}x_{h,i}y_{h,b}.
\end{equation}
There is at most one $b$ in this sum.

\begin{proposition}[Matrix and interpolation criterion]\label{prop:interpolation}
The downloaded coefficient conditions are unisolvent on the full
order-$(t-1)$ space if and only if $\mathcal E_H$ is invertible. In
the coefficient bases above,
\begin{equation}\label{eq:transposecriterion}
\mathcal E_H=\A_H^{\mathsf T}.
\end{equation}
\end{proposition}
\begin{proof}
The coefficient with remaining index $\gamma$ in a restricted $B_c$ is
$\sum_{i,b}x_{h,i}y_{h,b}c_{i,\gamma+e_b}$.
The column of $\A_H$ indexed by $(h,\gamma)$ is exactly the coefficient
column of $x_h\otimes y_hz^\gamma$. Comparing entries proves
\eqref{eq:transposecriterion}. A square linear coefficient system is
unisolvent precisely when its kernel is zero.
\end{proof}
To execute repair directly in monomial coordinates, one solves
$\mathcal E_Hq_f=(d_{h,\gamma})$ and outputs
$c_{f,\delta}=\sum_i x_{f,i}(q_f)_{i,\delta}$.

\subsection{Reconstruction and all-node exact repair}
The following interpolation formula will serve both the data collector
and the replacement node. Write $\binom Ks$ for the $s$-subsets of $K$.
\begin{lemma}[Interpolation on subsets]\label{lem:subsetinterpolation}
Let $s\geq1$ and let $K$ index $m+s-1$ vectors $y_h\in F^m$,
any $m$ of which are independent. For $I\in\binom Ks$, choose a
nonzero linear form $\ell_I$ vanishing on the $m-1$ vectors indexed by
$K\setminus I$. Put
\[
L_I(Z_1,\ldots,Z_s)=
\frac{\prod_{j=1}^s\ell_I(Z_j)}{\prod_{i\in I}\ell_I(y_i)}.
\]
Every polynomial $P$ that is linear in each vector variable and
invariant under permutations of the $s$ variables satisfies
\begin{equation}\label{eq:subsetinterpolation}
P(Z_1,\ldots,Z_s)=\sum_{I\in\binom Ks} P(y_I)L_I(Z_1,\ldots,Z_s),
\end{equation}
where $y_I$ lists the vectors in $I$ in any order.
\end{lemma}
\begin{proof}
Independence gives $\ell_I(y_i)\ne0$ for $i\in I$. Thus $L_I(y_I)=1$
and $L_I(y_J)=0$ for every other $s$-subset $J$. These polynomials
are linearly independent. Their number $\binom{m+s-1}{s}$ equals the
dimension of the space, since one coefficient can be chosen freely
for each nondecreasing $s$-tuple from $\{0,\ldots,m-1\}$.
\end{proof}
For $y_h=y(a_h)$ with $y(T)=(1,T,\ldots,T^{m-1})$, one may take
$\ell_I(y(T))=\prod_{h\in K\setminus I}(T-a_h)$.
Substituting $Z_j=y(T_j)$ then gives the ordinary Lagrange formula
on unordered subsets for symmetric polynomials of degree less than
$m$ in each scalar variable~\cite[Th.~1.1]{krickszanto2026}.
This substitution preserves all
coefficients in \eqref{eq:polarpoly}.

\begin{proposition}[Polynomial coefficient code]\label{prop:code}
Assume that any $t$ first-factor vectors $x_h$ are independent, any $m$
second-factor vectors $y_h$ are independent, and $\mathcal E_H$ is
invertible for every $d$-element helper set. Then
\eqref{eq:nodepoly}--\eqref{eq:downloadpoly} give a linear all-node
exact-repair minimum storage regenerating code with $d=tr$,
$k=(t-1)r+1$, and \eqref{eq:alphabeta} for $n\geq d+1$.
\end{proposition}
\begin{proof}
Fix $k=m+t-1$ available nodes, with index set $K$. For any $t$ distinct
indices $h_1,\ldots,h_t$, each of these nodes evaluates its stored
polynomial at the other $t-1$ second-factor vectors. The resulting $t$
values are
$B_a(x_{h_j};y_{h_1},\ldots,y_{h_t})$, $1\leq j\leq t$.
The $x_{h_j}$ form a basis, so they determine
$B_a(U;y_{h_1},\ldots,y_{h_t})$ for arbitrary $U$.

These are the values in \eqref{eq:subsetinterpolation} with $s=t$.
Applying the formula for each $U=e_i$ recovers all file coefficients.

For repair, \eqref{eq:downloadpoly} supplies exactly $\mathcal E_Hc_f$,
where $c_f$ is the coefficient array of $Q_f$.
Invertibility determines $Q_f$, and substituting $x_f$ recovers $C_f$.
There are $\beta$ transmitted symbols per helper and
$\beta=\alpha/(d-k+1)$, as required.
\end{proof}

\subsection{Repair by subset interpolation}
For $t=3$ the unknown repair polynomial $Q_f$ has two vector variables
besides $U$, and for $t=4$ it has three. Each interpolation value
$Q_f(U;y_I)$ is a linear form in $U$. The helpers indexed by $I$
give all but one of its coordinates in a suitable basis.

\begin{proposition}[Reduced repair system]\label{prop:subsetrepair}
Assume the two node independence conditions of Proposition~\ref{prop:code}
and let $t\geq3$. Put $s=t-1$. In any $d$-element helper set $H$,
choose $K$ of size $k-1=m+s-1$, and remove one index from $K$ to
obtain $K'$. For $I=\{i_1<\cdots<i_s\}\subset K$ set
$v_I(U)=\det[x_{i_1}\ \cdots\ x_{i_s}\ U]$ and choose $\ell_I$ as
in Lemma~\ref{lem:subsetinterpolation}. Define $R$ with columns
$I\in\binom Ks$ and rows $(h,J)$ with $h\in H\setminus K$,
$J\in\binom{K'}{s-1}$. Its entries are zero unless $J\subset I$.
For $I=J\cup\{i\}$ set
\begin{equation}\label{eq:subsetrepair}
 R_{(h,J),I}=v_I(x_h)\frac{\ell_I(y_h)}{\ell_I(y_i)}.
\end{equation}
Then $R$ is square of order $\alpha$ and
$\dim\ker R=\dim\ker\mathcal E_H$.
\end{proposition}
\begin{proof}
Let $Q$ be an unknown repair polynomial. By
\eqref{eq:subsetinterpolation} its coordinates are the linear forms
$q_I(U)=Q(U;y_I)$. For each $i\in I$, the slice supplied by helper
$i$ gives $q_I(x_i)$ by evaluation at $y_{I\setminus\{i\}}$.
Conversely these values give the entire slice from helper $i$ by the
lemma on $K\setminus\{i\}$ with $s-1$ variables. For zero data from
$K$, the form $q_I$ therefore vanishes on the $s=t-1$ independent
vectors $x_i$, $i\in I$. Equivalently $q_I=b_Iv_I$ for a unique
scalar $b_I$. There is one free scalar per $I$.

For each remaining helper $h$, the lemma on $K'$ identifies its slice
with the values $Q(x_h;y_h,y_J)$. In the interpolation formula for
this value, $L_I(y_h,y_J)$ vanishes unless $J\subset I$. If
$I=J\cup\{i\}$, cancellation gives $\ell_I(y_h)/\ell_I(y_i)$.
Hence the remaining homogeneous equations are exactly $Rb=0$.
This correspondence is bijective. There are $\binom{k-1}{t-1}=\alpha$
columns and $(d-k+1)\binom{k-2}{t-2}=r\beta=\alpha$ rows.
\end{proof}
For actual downloads, interpolate one particular form $q_I^0$
satisfying the $s$ known values from $I$. Write $q_I=q_I^0+b_Iv_I$
and subtract the contribution of the $q_I^0$ from the remaining
observations. Solving \eqref{eq:subsetrepair} for $b$ and applying
\eqref{eq:subsetinterpolation} at $U=x_f$ yields the failed node.
For example, at $r=2$, the reduced system $R$ has order $6$ for $t=3$
and $20$ for $t=4$, compared with $18$ and $80$ in
\eqref{eq:transposecriterion}. Once the node coordinates are fixed,
repair decoding uses $R$; to prove determinant nonvanishing we work
with $\A_H=\mathcal E_H^{\mathsf T}$, whose polynomial entries remain
well-defined along singular degenerations of the node configuration.

\subsection{Example: The $t=2$ product-matrix endpoint}\label{subsec:t2pm}
At $t=2$, setting $d=tr$, $k=(t-1)r+1$, and $m=(t-1)(r-1)+1$ in
\eqref{eq:alphabeta} gives
$(k,d,m,\alpha,\beta)=(r+1,2r,r,r,1)$, matching the product-matrix MSR
code of Rashmi et~al.~\cite{pm} (see also~\cite[Sec.~3]{eidduursma2016},
\cite[Sec.~V-A]{patrabarg2025}, \cite{teamPatent2025}). An order-two file
array consists of two $r\times r$ symmetric matrices $S_0,S_1$ carrying
$2\binom{r+1}{2}=r(r+1)=k\alpha$ independent symbols, with polarized
bivariate form
\[
B_a\bigl(U;Z^{(1)},Z^{(2)}\bigr)
=U_0\bigl(Z^{(1)}\bigr)^{\mathsf T}S_0Z^{(2)}
+U_1\bigl(Z^{(1)}\bigr)^{\mathsf T}S_1Z^{(2)}.
\]
Choosing $x_h=(1,\xi_h)^{\mathsf T}$ and
$y_h=(1,a_h,\ldots,a_h^{r-1})^{\mathsf T}$, node $h$ stores the
coefficients of $C_h(Z^{(2)})=y_h^{\mathsf T}(S_0+\xi_h S_1)Z^{(2)}$,
namely the $r$-symbol row $y_h^{\mathsf T}(S_0+\xi_h S_1)$ of~\cite{pm}.

To repair node $f$ from a $2r$-helper set $H$, helper $h\in H$ evaluates
$Z^{(2)}=y_f$ and transmits the scalar
$d_{h\to f}=y_h^{\mathsf T}(S_0y_f+\xi_h S_1y_f)$.
Setting $\xi_h=a_h^r$ with distinct $a_h$ and distinct $a_h^r$ (which is
achievable over any finite field with $q>r(n-1)$, since each fiber of
$a\mapsto a^r$ has size at most $r$ and hence at least
$\lceil q/r\rceil\geq n$ image values exist) turns $d_{h\to f}=G_f(a_h)$
into point evaluations of the univariate polynomial
\begin{equation}\label{eq:t2repairpoly}
G_f(Z)=\sum_{j=0}^{r-1}(S_0y_f)_jZ^j+\sum_{j=0}^{r-1}(S_1y_f)_jZ^{r+j}
\end{equation}
of degree less than $2r$. Simultaneously, $\mathcal E_H=\A_H^{\mathsf T}$
reduces to the invertible $2r\times 2r$ Vandermonde matrix
$V_{2r}=(a_h^b)_{h\in H,\,0\leq b<2r}$. Interpolating $G_f$ in $O(d^2)$
field operations (or $O(M(d)\log d)$ with fast polynomial arithmetic~\cite{vzgg})
recovers $u=S_0y_f$ and $v=S_1y_f$, yielding the failed node
$c_f^{\mathsf T}=u^{\mathsf T}+a_f^r v^{\mathsf T}$ in $O(r)$ further
operations.

From a degeneration viewpoint, splitting $H$ into two $r$-node groups
with $x_h=e_0$ and $x_h=e_1$ already makes $\A_H$ block-diagonal with two
invertible $r\times r$ MDS blocks, because $\dim Y=r$. For $t\geq3$, by
contrast, $m-r=(t-2)(r-1)>0$: an $r$-node group spans a proper subspace
of $Y$, creating a singular baseline whose kernel and cokernel must be
matched by first-order perturbation.

\section{The Perturbation--Syzygy Framework}\label{sec:structure}
All determinant proofs in this paper follow a common three-step reduction.
First, we degenerate the repair map $\A_H$ along $X$-coordinate channels
(preceded, for $t=4$ and $r\geq3$, by Hermite clustering of Vandermonde
nodes in $Y$) so that the baseline operator decomposes into homogeneous
polynomial multiplication maps. Second, we identify each block's kernel
by Koszul syzygies and its cokernel by a quotient symmetric power. Third,
we apply first-order Schur lifting to certify that the perturbation
supplies all missing interpolation conditions. Table~\ref{tab:architecture}
summarizes how each parameter regime instantiates this framework.

\begin{roadmaptable}
\centering\small\setlength{\tabcolsep}{5pt}
\caption{Unified degeneration, Koszul-syzygy, and Schur-lifting architecture across code families.}
\label{tab:architecture}
\begin{tabular}{p{0.15\textwidth}p{0.23\textwidth}p{0.27\textwidth}p{0.25\textwidth}}
\toprule
Family & Baseline degeneration & Koszul kernel \& quotient cokernel & First-order Schur lift \\
\midrule
$t=2$, $r\geq1$ & Two $r$-node $X$-channels & Zero defect ($\dim(Y/W_i)=0$) & None needed ($\A_0$ invertible) \\
$t=3$, $r\geq2$ & Three $r$-node $X$-channels, fixed MDS $y_h$ & $r$-form syzygies $\kappa_{ab}$; $\Sym^2(Y/W_i)$ of dim $\binom r2$ & Cyclic $X$-pencil; frame-product minors (Lemma~\ref{lem:frame}) \\
$(t,r)=(4,2)$ & Four $2$-node $X$-channels, fixed MDS $y_h$ & $2$-form syzygies $(L_i\ell_i,-F_i\ell_i)$; $\Sym^3(Y/W_i)$ of dim $4$ & $X$-pencil; binary Sylvester resultant (Proposition~\ref{prop:t4r2}) \\
$t=4$, $r\geq3$ & Vandermonde Hermite $\eps$-clusters + $X$-channels & Support-triangular blocks ($|S|\leq3$) via $2$- and $3$-form syzygies & Local $\eta$-lifts + pure-support $\eps$-decoupling (Proposition~\ref{prop:globalassembly}) \\
$(t,r)=(5,2),(6,2)$ & Five/six $2$-node $X$-channels, fixed MDS $y_h$ & $2$-form syzygies; $\Sym^{t-1}(Y/W_i)$ of dim $15,56$ & Cyclic $X$-pencil; syzygy gluing \& Laurent/Schur lifts (Theorem~\ref{thm:higherrepair}, Appendices~\ref{app:t5}--\ref{app:t6}) \\
\bottomrule
\end{tabular}
\end{roadmaptable}

\subsection{Multiplication maps and Koszul syzygies}
For linearly independent forms $\ell_1,\ldots,\ell_s\in Y$ with span
$W=\Span(\ell_1,\ldots,\ell_s)$, consider the homogeneous multiplication map
\begin{equation}\label{eq:multblock}
\mu_W:(\Sym^jY)^s\longrightarrow\Sym^{j+1}Y,\qquad
(q_a)_{a=1}^s\longmapsto\sum_{a=1}^s\ell_a q_a.
\end{equation}

\begin{lemma}[Multiplication cokernels and Koszul syzygies]\label{lem:koszul}
Let $\ell_1,\ldots,\ell_s\in Y$ be linearly independent over an arbitrary
field $F$, with $W=\Span(\ell_1,\ldots,\ell_s)$.
\begin{enumerate}
\item \emph{Quotient cokernel:} For every $j\geq0$, the image of $\mu_W$
is the degree-$(j+1)$ component of the ideal generated by $W$, and
reduction modulo $W$ induces $\coker\mu_W\simeq\Sym^{j+1}(Y/W)$.
\item \emph{Degree-one $r$-form syzygies:} If $j=1$, $s=r$, and
$\dim Y=2r-1$, then $\mu_W:Y^r\to\Sym^2Y$ is square of order $r(2r-1)$
and corank $\nu=\binom r2$. A right null basis is given by
$\kappa_{ab}=\ell_be_a-\ell_ae_b$ for $1\leq a<b\leq r$.
\item \emph{Two- and three-form syzygies:}
For $s=2$ and any $j\geq1$, $\ell_1q_1+\ell_2q_2=0$ in $\Sym^{j+1}Y$ if
and only if $(q_1,q_2)=(\ell_2h,-\ell_1h)$ for a unique
$h\in\Sym^{j-1}Y$. For $s=3$ and $j=2$,
$\ell_1q_1+\ell_2q_2+\ell_3q_3=0$ in $\Sym^3Y$ if and only if
\begin{equation}\label{eq:koszul3}
\begin{aligned}
q_1&=\ell_2\tau_{12}+\ell_3\tau_{13},\qquad
q_2=-\ell_1\tau_{12}+\ell_3\tau_{23},\\
q_3&=-\ell_1\tau_{13}-\ell_2\tau_{23}
\end{aligned}
\end{equation}
for $(\tau_{12},\tau_{13},\tau_{23})\in Y^3$, unique modulo the one-dimensional
subspace $F\cdot(\ell_3,-\ell_2,\ell_1)$.
\end{enumerate}
\end{lemma}
\begin{proof}
Complete $\ell_1,\ldots,\ell_s$ to a basis of $Y$, identifying $\Sym Y$
with a polynomial ring in which $\ell_1,\ldots,\ell_s$ are distinct
indeterminates. Monomials of degree $j+1$ divisible by at least one
$\ell_a$ span $\operatorname{im}\mu_W$, while the remaining monomials
project to a basis of $\Sym^{j+1}(Y/W)$, proving~1).

In~2), source and target both have dimension $r(2r-1)$ and
$\dim(Y/W)=r-1$, so $\dim\ker\mu_W=\dim\Sym^2(Y/W)=\binom r2$. Each
$\kappa_{ab}$ belongs to $\ker\mu_W$, and the $\ell_b$ coordinate in
component $a$ establishes linear independence.

In~3), $\ell_1$ and $\ell_2$ are coprime indeterminates, so
$\ell_1q_1=-\ell_2q_2$ holds if and only if $q_1=\ell_2h$ and
$q_2=-\ell_1h$ for a unique $h\in\Sym^{j-1}Y$. For $s=3$, $\ell_3$ is a
nonzero divisor in $\Sym(Y/\Span(\ell_1,\ell_2))$, so reducing
$\ell_1q_1+\ell_2q_2+\ell_3q_3=0$ modulo $(\ell_1,\ell_2)$ yields
$q_3=-\ell_1\tau_{13}-\ell_2\tau_{23}$ for some $\tau_{13},\tau_{23}\in Y$.
Then $\ell_1(q_1-\ell_3\tau_{13})+\ell_2(q_2-\ell_3\tau_{23})=0$, and the
$s=2$ case supplies $\tau_{12}\in Y$. If $(q_1,q_2,q_3)=0$, then
$\ell_1\tau_{13}+\ell_2\tau_{23}=0$ gives
$(\tau_{13},\tau_{23})=(-c\ell_2,c\ell_1)$ with $c\in F$, and
$\ell_2\tau_{12}+\ell_3\tau_{13}=0$ forces $\tau_{12}=c\ell_3$.
\end{proof}

\subsection{First-order Schur lifting and support-triangular elimination}
\begin{lemma}[First-order Schur lifting]\label{lem:lift}
Let $M(s)=M_0+sM_1+O(s^2)$ be an $N$-square matrix over $F[[s]]$,
and put $\nu=\dim\ker M_0$. If $E:\ker M_0\to\coker M_0$ induced by
$M_1$ is invertible, then
\begin{equation}\label{eq:lift}
\det M(s)=\gamma_0 s^\nu\det E+O(s^{\nu+1}),\quad\gamma_0\ne0.
\end{equation}
Conversely, a nonzero $s^\nu$ coefficient implies that $E$ is invertible.
More generally, if $M(s)$ has entries in $\mathcal O[[s]]$ for a
discrete valuation ring $\mathcal O$ and $M_0$ reduces via invertible
row and column operations over $\mathcal O$ to
$\operatorname{diag}(I_{N-\nu},0)$, then \eqref{eq:lift} holds over
$\mathcal O[[s]]$ with $\gamma_0\in\mathcal O^\times$ a unit, so
$[s^\nu]\det M(s)$ has the same valuation as $\det E$.
\end{lemma}
\begin{proof}
Constant row and column operations put $M_0$ into
$\operatorname{diag}(I_{N-\nu},0)$. In these bases,
\[
M(s)=\begin{pmatrix}
I+sB+O(s^2)&sC+O(s^2)\\sD+O(s^2)&sE+O(s^2)
\end{pmatrix}.
\]
The upper-left block is invertible over $F[[s]]$ (and over
$\mathcal O[[s]]$ in the discrete valuation ring setting), with Schur
complement $sE+O(s^2)$. Taking determinants proves \eqref{eq:lift} up to
the determinant $\gamma_0$ of the constant basis changes (nonzero over $F$,
and a unit in $\mathcal O^\times$ when the basis changes are invertible
over $\mathcal O$), gives valuation at least $\nu$ in general, and
yields the converse.
\end{proof}
In coordinates, $E=L_0M_1N_0$, where the columns of $N_0$ and rows of
$L_0$ form right and left null bases of $M_0$. Any variation of the form
$LM_0+M_0R$ induces the zero map from $\ker M_0$ to $\coker M_0$.

\begin{lemma}[Specialized-coefficient lifting]\label{lem:coefficientlift}
Let $R$ be an integral domain, $L$ its fraction field, and
$M(s)\in\operatorname{Mat}_N(R[[s]])$. Suppose the corank of $M(0)$
over $L$ is at least $\delta$, and put $D_\delta=[s^\delta]\det M(s)\in R$.
If a coefficientwise specialization $\sigma:R\to K[[\eta]]$ satisfies
$[\eta^u]\sigma(D_\delta)\ne0$ for some $u$, then over $L$ both the
zero-order corank and $\ord_s\det M(s)$ equal $\delta$, and the map
induced by $[s]M(s)$ from $\ker M(0)$ to $\coker M(0)$ is an isomorphism.
The same holds over localizations of $R$ at elements mapped to units by
$\sigma$.
\end{lemma}
\begin{proof}
Nonvanishing of $[\eta^u]\sigma(D_\delta)$ implies $D_\delta\ne0$ in $R$.
If $\nu$ denotes the corank of $M(0)$ over $L$, Lemma~\ref{lem:lift} gives
$\delta\leq\nu\leq\ord_s\det M(s)\leq\delta$, so equality holds throughout
and the converse of Lemma~\ref{lem:lift} applies.
\end{proof}

\begin{lemma}[Support-triangular elimination]\label{lem:supportelimination}
Let $\mathcal G$ be a finite ground set and put
$\mathcal S=\{S\subseteq\mathcal G:|S|\leq s_0\}$.
Let $V=\bigoplus_{S\in\mathcal S}V_S$ and
$W=\bigoplus_{S\in\mathcal S}W_S$, with $\dim V_S=\dim W_S$.
Suppose $A_{T,S}=0$ unless $S\subseteq T$, and write $B_S=A_{S,S}$.
If $B_\varnothing$ and every $B_S$ with $|S|\geq2$ are invertible,
then restriction to singleton components gives an isomorphism
\[
\ker A\simeq\bigoplus_{g\in\mathcal G}\ker B_{\{g\}},
\]
and the extension of a vector supported on $\{g\}$ vanishes at all
labels $T$ with $g\notin T$.
\end{lemma}
\begin{proof}
The empty row forces $q_\varnothing=0$, after which each singleton row
reads $B_{\{g\}}q_{\{g\}}=0$ independently. Given arbitrary
$q_{\{g\}}\in\ker B_{\{g\}}$, define recursively in increasing $|T|\geq2$
\begin{equation}\label{eq:supportrecursion}
q_T=-B_T^{-1}\sum_{S\subsetneq T}A_{T,S}q_S.
\end{equation}
Because $A_{T',T}=0$ unless $T\subseteq T'$, setting $q_T$ leaves all
smaller or incomparable rows unchanged and uniquely solves row $T$.
Induction on $|T|$ in \eqref{eq:supportrecursion} shows that an input
supported on $\{g\}$ has $q_T=0$ whenever $g\notin T$.
\end{proof}

\section{The $t=3$ Determinant for Fixed MDS Nodes}\label{sec:t3}
Here $m=2r-1$, $\alpha=r(2r-1)$, and $\beta=m$.
An order-two array in \eqref{eq:polarpoly} is a symmetric matrix $S_i$
for each $X$ coordinate. The observations at helper $h$ are exactly
\begin{equation}\label{eq:t3matrixdata}
d_h=\left(\sum_{i=0}^2x_{h,i}S_i\right)y_h.
\end{equation}
Unisolvence means that these direction observations uniquely determine
all three symmetric matrices. We prove it by a nonzero determinant
coefficient rather than by counting observations.

\begin{theorem}[Fixed-MDS degree-two unisolvence]\label{thm:t3}
For any field $F$, let $r\geq2$ and fix $3r$ vectors
$y_h\in F^{2r-1}$ such that every $2r-1$ are independent.
The determinant of \eqref{eq:t3matrixdata}, as a square coefficient
system, is a nonzero polynomial in the $x_h\in F^3$.
It remains nonzero after setting every first $x$ coordinate to one.
\end{theorem}

\begin{lemma}[Frame-product determinant]\label{lem:frame}
Let $u=r-1$ and let $v_1,\ldots,v_r\in F^u$ have every $u$ columns
independent. Let $\mathcal B(v)$ have columns the ordinary quadratic
coefficient columns of $v_av_b$, $a<b$. It is $\binom r2$-square, and
\begin{equation}\label{eq:frameminorproduct}
\det\mathcal B(v)=\pm\prod_{a=1}^r
\det[v_1,\ldots,\widehat v_a,\ldots,v_r]\ne0.
\end{equation}
\end{lemma}
\begin{proof}
Put $U=[v_1\ \cdots\ v_u]$ and write $v_r=\sum_{a=1}^uc_av_a$;
all $c_a$ are nonzero by the independence assumption. In coordinates with
$U=I_u$, the columns $v_av_b$, $a<b<r$, give the off-diagonal monomials,
and subtracting them from $v_av_r$ leaves $c_av_a^2$.
The determinant in these coordinates is therefore $\pm\prod_ac_a$.

The ordinary quadratic coefficient change induced by $U$ has
determinant $(\det U)^{u+1}$: this holds for diagonal scalings by
exponent counting, for elementary additions by triangularity, and
for coordinate swaps by counting transposed monomials.
In the original coordinates the determinant is therefore
$\pm(\det U)^{u+1}\prod_ac_a$. Writing each $c_a$ via Cramer's rule as a
replacement minor divided by $\det U$ and canceling $u$ factors of
$\det U$ yields \eqref{eq:frameminorproduct}.
\end{proof}

\begin{proof}[Proof of Theorem~\ref{thm:t3}]
Partition the $3r$ helpers into three groups $G_i$ of size $r$, indexed
modulo three, and define the group subspaces
$W_i=\Span(y_{i,1},\ldots,y_{i,r})$.
Over an auxiliary extension choose distinct $\lambda_1,\ldots,\lambda_r$
and specialize
\begin{equation}\label{eq:t3pencil}
x_{i,a}(s)=e_i+s\lambda_ae_{i+1}.
\end{equation}
At $s=0$, the repair map is $\operatorname{diag}(\mu_{W_0},\mu_{W_1},\mu_{W_2})$.
By Lemma~\ref{lem:koszul}, each block has corank $\nu=\binom r2$, Koszul
null basis $\kappa_{ab}$ ($a<b$), and cokernel $\Sym^2(Y/W_i)$.

Choose any full-row-rank projection $P_{i+1}:Y\to F^{r-1}$ with kernel
$W_{i+1}$, and set $v_{i,a}=P_{i+1}y_{i,a}$.
Every $r-1$ of these columns are independent, since any relation lifts
modulo $W_{i+1}$ to a relation among $2r-1$ original MDS vectors. The
first-order image of $\kappa_{ab}$ from group $i$ in $\Sym^2(Y/W_{i+1})$ is
$(\lambda_a-\lambda_b)v_{i,a}v_{i,b}$, giving the $3\nu$-square induced map
\begin{equation}\label{eq:t3cyclicmatrix}
E_3=\begin{pmatrix}0&0&T_2\\T_0&0&0\\0&T_1&0\end{pmatrix},
\end{equation}
where
\[
T_i=\mathcal B(v_{i,1},\ldots,v_{i,r})
\operatorname{diag}_{a<b}(\lambda_a-\lambda_b).
\]
Lemma~\ref{lem:frame} yields
\begin{equation}\label{eq:t3edgecertificate}
\begin{split}
\det T_i={}&\pm\prod_{a<b}(\lambda_a-\lambda_b)\\
&\quad\cdot\prod_a
\det[v_{i,1},\ldots,\widehat v_{i,a},\ldots,v_{i,r}]\ne0.
\end{split}
\end{equation}
Hence $E_3$ is invertible, and Lemma~\ref{lem:lift} gives
\begin{equation}\label{eq:t3leadingcertificate}
\det\A(s)=\gamma_0 s^{3\binom r2}\prod_i\det T_i
+O\bigl(s^{3\binom r2+1}\bigr),\qquad\gamma_0\ne0.
\end{equation}
Thus $\det\A$ is a nonzero polynomial over the extension and hence over
$F$. Finally, $\det\A$ is homogeneous of degree $\beta=m$ in each $x_h$,
so setting $x_{h,0}=1$ preserves distinct monomials and maintains
nonvanishing.
\end{proof}

\section{The $t=4$ Determinant for Vandermonde Nodes}\label{sec:t4}

Each column of $\A=\mathcal E_H^{\mathsf T}$ multiplies a quadratic
monomial by a node linear form. We bring the Vandermonde nodes
together in groups of four and compute which cubic coefficient
conditions are lost in the limit. Moving the nodes apart restores
those conditions to first order. Throughout this section,
$m=3r-2$, $\alpha=\binom{3r}{3}$ and $\beta=\binom{3r-1}{2}$.
\begin{theorem}[Joint Vandermonde nonvanishing]\label{thm:t4}
For every $r\geq2$, the determinant of
\[
\begin{aligned}
\A&:\bigoplus_{h=1}^{4r}\Sym^2Y\longrightarrow F^4\otimes\Sym^3Y,\\
&(q_h)\longmapsto\sum_hx_h\otimes y(a_h)q_h,
\end{aligned}
\]
where $y(a)=(1,a,\ldots,a^{3r-3})$, is a nonzero polynomial over every
field $F$ in the joint variables $(x_h,a_h)$. Setting each first $x$ coordinate
equal to one preserves this conclusion.
\end{theorem}

\noindent\emph{Running case: $r=3$.}
Then $m=7$, $d=12$, $\alpha=84$, $\beta=28$, and the repair matrix
has order $336$. There are two local groups. Its zero-order support
blocks have orders
$4+2\cdot76+180=336$, corresponding to the empty support,
two singletons, and one pair. We follow this case through the local defect calculation,
the correction equation, and the final eight-dimensional lift below.
For general $r$, Lemmas~\ref{lem:single4}--\ref{lem:triple4} treat one,
two, and three groups, respectively. Only the single-group blocks have
a kernel. Proposition~\ref{prop:globalassembly} shows that the actual
node displacement $\eps$ resolves their four missing conditions
independently in each group. The parameters $\eta,\lambda,\mu$ enter
only the local first-factor calculations.

\subsection{Hermite degeneration and support ordering}\label{subsec:degeneration}
Write $G=r-1$ and let $K$ be any extension field of $F$ containing $G+4$
pairwise distinct elements $A_0,\ldots,A_3,a_1,\ldots,a_G$ and four
pairwise distinct elements $b_0,\ldots,b_3$: to prove
Theorem~\ref{thm:t4} we take $K$ to be the rational function field over
$F$ in these parameters and descend polynomial nonvanishing to $F$ at the
end, whereas Theorem~\ref{thm:interpolatednodes} and
Appendix~\ref{app:selector} take $K$ directly to be a finite field
containing such distinct elements (since every local proof below uses
only pairwise distinctness of $A_0,\ldots,A_3,a_1,\ldots,a_G$, pairwise
distinctness of $b_0,\ldots,b_3$, and $\lambda\neq\mu$ with
$\lambda,\mu\notin\{0,1\}$, never algebraic independence). We extend $X$
and $Y$ to $K$ without changing their notation during the block
calculations. Put
\begin{align}\label{eq:localforms}
Y&=Kf\oplus\bigoplus_{g=1}^G Z_g,&
Z_g&=\Span(z_{g,0},z_{g,1},w_g),\\
c_{g,i}&=(A_i-a_g)^{-1},&
L_{g,i}&=z_{g,0}+b_iz_{g,1}+b_i^2w_g.\nonumber
\end{align}
Any three distinct $L_{g,i}$ are independent. Also $w_g$ and any two
distinct $L_{g,i}$ are independent, since reduction modulo $w_g$ leaves
the two independent forms $z_{g,0}+b_iz_{g,1}$.
Within each four-node group, indices $i,j,k$ are taken modulo four.
The index $g$ always denotes a local group; the index $0$ in $y_{0,i}$
denotes a central node, not a local group.

Use ordinary nodes $A_i$ and $a_g+\eps b_i$. An invertible coordinate
change over $K((\eps))$ and nonzero individual node scalings give
\begin{align}\label{eq:jets}
y_{0,i}&=f+\sum_gc_{g,i}w_g
 +\eps\sum_g(c_{g,i}^2-S_gc_{g,i})z_{g,1}+O(\eps^2),\\
y_{g,i}&=L_{g,i}+\eps b_i^3
 \left(f+S_gw_g+\sum_{h\neq g}\frac{w_h}{a_g-a_h}\right)+O(\eps^2),\nonumber\\
S_g&=3\sum_{h\neq g}(a_g-a_h)^{-1}.\nonumber
\end{align}
Appendix~\ref{app:hermite} derives \eqref{eq:jets} over an arbitrary
field by polynomial interpolation; these invertible coordinate changes
and node scalings over $K((\eps))$ preserve determinant nonvanishing.

At $\eps=0$, label a central source monomial $q$ by the groups it uses,
and a source monomial at local group $g$ by $\supp(q)\cup\{g\}$.
Here $\supp(q)$ is the set of $h$ for which $q$ contains a variable in
$Z_h$; powers of $f$ do not add a group to the support.
The target support always contains the source label. Ordering supports by
inclusion makes $\A_0$ block triangular. Table~\ref{tab:blocks} gives the
square diagonal blocks. There are no larger supports because the target
degree is three.
For example, a central source $fz_{g,0}$ produces a term with support
$\{g\}$ when multiplied by $f$ or $w_g$, and terms with support
$\{g,h\}$ when multiplied by $w_h$ for $h\neq g$.
This illustrates both the diagonal blocks and the off-diagonal terms
that must be handled in the assembly step.
\begin{roadmaptable}
\centering
\caption{Support blocks of the zero-order $t=4$ repair map.}
\label{tab:blocks}
\begin{tabular}{clrr}
\toprule
Support size & Target monomials before tensoring with $X$ & Count & Block order\\
\midrule
0 & $f^3$ &1&4\\
1 & $f^2Z_g,\ f\Sym^2Z_g,\ \Sym^3Z_g$ & $3+6+10$&76\\
2 & $fZ_gZ_h,\ \Sym^2Z_gZ_h,\ Z_g\Sym^2Z_h$ &$9+18+18$&180\\
3 & $Z_gZ_hZ_\ell$ &27&108\\
\bottomrule
\end{tabular}
\end{roadmaptable}
The source counts can be checked without constructing a matrix.
For a single support, each central node contributes the nine quadratics
$fZ_g\oplus\Sym^2Z_g$, while each local node contributes
$\Sym^2(Kf\oplus Z_g)$ of dimension ten: $4\cdot9+4\cdot10=76$.
For a double support, a central node contributes $Z_gZ_h$ of dimension
nine; a node in group $g$ contributes
$fZ_h\oplus Z_gZ_h\oplus\Sym^2Z_h$, of dimension $3+9+6=18$.
Thus $4\cdot9+2\cdot4\cdot18=180$.
For a triple support, each of its twelve local nodes contributes a
nine-dimensional product of the other two groups, giving $108$.
The empty block has four sources, namely $f^2$ at each central
node, and four targets, namely $X\otimes f^3$; its matrix is that of
the four central $x$ vectors. There are no central sources with three
groups because a quadratic uses at most two, while a local source adds
only its own group to that support. Consequently the table lists every
source and target type. As a global dimension check,
\[
4+76G+180\binom G2+108\binom G3
=4\binom{3G+3}{3}=4\alpha.
\]
Here binomial coefficients count actual subsets and vanish when the
support size exceeds $G$. Thus $s_0=3$ in
Lemma~\ref{lem:supportelimination} covers all rows, including the empty block.

\subsection{Single support and fixed-MDS unisolvence at $(t,r)=(4,2)$}
Each $76$-square singleton block leaves four missing conditions; adjoining
the invertible four-dimensional empty block gives an $80$-square
one-group repair map. An auxiliary $X$-channel perturbation in $\eta$
isolates a mixed determinant coefficient, after which
Lemma~\ref{lem:coefficientlift} identifies both the generic corank and
the $\eps$ lift.

\begin{lemma}[The single-support defect]\label{lem:single4}
For generic central and local $x$ vectors, the $76$-square single-support
block has corank four. In the full one-group $80$-square repair map,
the induced $\eps$ map from kernel to cokernel is invertible.
\end{lemma}
\begin{proof}
Write $c_i=c_{1,i}$, $a=a_1$, omit the group subscript, and set
$R=K[x_{0,i,j},x_{1,i,j}:0\leq i,j<4]$ and $L=\operatorname{Frac}R$.
The normalized one-group matrix lies in $\operatorname{Mat}_{80}(R[\eps])$,
with exact node forms
\begin{equation}\label{eq:singleexact}
F_i=f+c_iw+\eps c_i^2z_1+\eps^2c_i^3z_0,
\qquad L_i=z_0+b_iz_1+b_i^2w+\eps b_i^3f.
\end{equation}
At $\eps=0$, projecting $Y$ onto $\Span(z_0,z_1)$ by killing $f,w$ and
taking the local $x_i$ as a basis of $X$ sends the central forms to zero
and leaves local images divisible by $z_0+b_iz_1$. Quotienting by
$z_0+b_iz_1$ leaves one missing cubic coordinate in each of the four
components, so the corank over $L$ is at least four. By
Lemma~\ref{lem:coefficientlift}, it suffices to exhibit a specialization
with nonzero $[\eta^{16}\eps^4]$ determinant coefficient.

\emph{Channel baseline and Koszul syzygies.}
Set the central and local first factors equal to $e_i$ (indices modulo
four), splitting the $80$-square map into four $20$-square multiplication
maps $(p_i,q_i)\mapsto F_ip_i+L_iq_i$ over the discrete valuation ring
$\mathcal O=K[[\eps]]$. Because the $(f,z_0)$ minor of $(F_i,L_i)$ is
$1-\eps^3c_i^3b_i^3\in\mathcal O^\times$, $(F_i,L_i,z_1,w)$ is an
$\mathcal O$-basis of $Y_{\mathcal O}$, with $z_1,w$ projecting to a free
basis of $\mathcal B_i=Y_{\mathcal O}/\Span_{\mathcal O}(F_i,L_i)$. In
the corresponding monomial bases of $\Sym^2Y_{\mathcal O}$ and
$\Sym^3Y_{\mathcal O}$, $(p_i,q_i)\mapsto F_ip_i+L_iq_i$ has constant
integer entries, free rank-four Koszul kernel
$\{(L_i\ell_i,-F_i\ell_i):\ell_i\in Y_{\mathcal O}\}$, and free rank-four
cokernel $\Sym^3\mathcal B_i$, reducing via invertible $\mathcal O$-row
and column operations to $\operatorname{diag}(I_{16},0_4)$ (total defect
$16$).

\emph{Binary Sylvester resultant.}
Keep local $x_i=e_i$ and perturb the central first factors to
\begin{equation}\label{eq:singleeta}
x_{0,i}=e_i+\eta(e_{i+1}+e_{i+2}).
\end{equation}
The induced first-order map in $\eta$ sends $\ell_i\in Y$ to the images
of $F_iL_i\ell_i$ in $\Sym^3\mathcal B_{i+1}$ and $\Sym^3\mathcal B_{i+2}$.
Since the constant term of $\det[F_j,L_j,F_k,L_k]$ is
$\pm(c_j-c_k)(b_j-b_k)\ne0$ for $j\ne k$, the restriction
$Y\to\mathcal B_{i+1}\oplus\mathcal B_{i+2}$ is an isomorphism with unit
determinant in $\mathcal O$. Consequently, at target $j$ with incoming
sources $i=j-1$ and $k=j-2$, the induced map takes the Sylvester form
\[
(u,v)\longmapsto Q_{ij}u+Q_{kj}v,
\qquad Q_{ij}=[F_iL_i]\in\Sym^2\mathcal B_j,
\]
in independent variables $u,v\in\mathcal B_j$. By multiplicativity of the
binary resultant, its determinant is, up to sign,
\begin{equation}\label{eq:resultant}
[F_i,F_k][F_i,L_k][L_i,F_k][L_i,L_k],
\end{equation}
where $[P,Q]$ denotes the $2\times2$ determinant in $\mathcal B_j$, which
has the same $\eps$-valuation as $\det(F_j,L_j,P,Q)$.

In the basis $(z_0,z_1,w,f)$, put
$v(u)=(1,u/\eps,u^2/\eps^2,u^3/\eps^2)^{\mathsf T}$, so that
$L_s=v(\eps b_s)$ and $F_s=\eps^2c_s^3v(A_s-a)$. In a four-column minor
with $h$ central nodes, row scalings contribute $-5$, central column
scalings contribute $2h$, and local--local Vandermonde differences
contribute $\binom{4-h}{2}$, giving exact $\eps$-valuation
\[
-5+2h+\binom{4-h}{2}.
\]
In \eqref{eq:resultant} the counts $h=3,2,2,1$ give valuations
$1,0,0,0$, so each target contributes valuation one. Multiplying over the
four targets and applying the discrete valuation ring form of
Lemma~\ref{lem:lift} over $\mathcal O[[\eta]]$ shows that
$[\eta^{16}]\det M$ has $\eps$-valuation four, hence
$[\eta^{16}\eps^4]\det M\ne0$. Lemma~\ref{lem:coefficientlift} with
$\delta=4$ and $u=16$ completes the proof.
\end{proof}

At $r=2$ ($m=4$, $d=8$), the same binary Sylvester resultant applies
directly without Vandermonde degeneration, extending fixed-MDS
unisolvence from $t=3$ to $(t,r)=(4,2)$.

\begin{proposition}[Fixed-MDS unisolvence at $(t,r)=(4,2)$]\label{prop:t4r2}
Over any field $F$, fix eight vectors $y_h\in F^4$ such that every four
are independent. Then $\det\A$ is a nonzero polynomial in the first
factors $x_h\in F^4$, and remains nonzero after setting every first $x$
coordinate to one.
\end{proposition}
\begin{proof}
Partition the eight vectors into four pairs $(F_i,L_i)$, $0\leq i<4$
(indices modulo four), set $x_{L_i}=e_i$, and perturb $x_{F_i}$ as in
\eqref{eq:singleeta}. At $\eta=0$, Lemma~\ref{lem:koszul} gives four
$20$-square blocks with Koszul kernels $\{(L_i\ell_i,-F_i\ell_i):\ell_i\in Y\}$
and cokernels $\Sym^3\mathcal B_i$, where $\mathcal B_i=Y/\Span(F_i,L_i)$
has dimension two. Because every four original vectors are independent,
$\det[F_j,L_j,F_k,L_k]\ne0$ for $j\ne k$, so
$Y\to\mathcal B_{i+1}\oplus\mathcal B_{i+2}$ is an isomorphism. At target
$j$ with distinct incoming indices $i=j-1$ and $k=j-2$, each bracket in
\eqref{eq:resultant} is proportional to a determinant of four distinct
MDS vectors (such as $\det[F_j,L_j,F_i,F_k]\ne0$) and is therefore
nonzero. Lemma~\ref{lem:lift} gives $[\eta^{16}]\det\A(\eta)\ne0$, and
homogeneity of degree $\beta=10$ in each $x_h$ preserves nonvanishing at
$x_{h,0}=1$.
\end{proof}

\subsection{Double support: two six-dimensional edge images}
\begin{lemma}[Double-support nonvanishing]\label{lem:double4}
Every $180$-square double-support diagonal block has a determinant that
is a nonzero polynomial in its $x$ variables over the auxiliary parameter
field.
\end{lemma}
\begin{proof}
For any two distinct local groups $g,h$, put
\[
Y_{g,h}=Kf\oplus Z_g\oplus Z_h,\qquad\dim Y_{g,h}=7.
\]
All calculations in this proof take place in $Y_{g,h}$ and its quotients,
not in the full $Y$ of dimension $3r-2$.
To extract this double-support diagonal block, retain the $g,h$ components
of each central node. Components in other groups contribute only to
larger-support rows and belong to the later assembly step.
Thus this local calculation applies for every $r$ for which $g,h$ exist;
when $r=2$, there is no double-support block. Abbreviate
\[
\ell_{g,i}=L_{g,i},\qquad\ell_{h,i}=L_{h,i},\qquad
C_i=f+c_{g,i}w_g+c_{h,i}w_h.
\]
The symbols $\ell_{g,i},\ell_{h,i}$ denote linear forms;
$A_i,a_g,b_i$ remain scalar node parameters.
At a common $x$ basis $e_i$, the $180$-square block splits into four
$45$-square maps. Quotient by $\ell_{g,i},\ell_{h,i}$ first and put
$U_i=Z_g/\Span(\ell_{g,i})$, $V_i=Z_h/\Span(\ell_{h,i})$.
The target becomes $fU_iV_i\oplus\Sym^2U_iV_i\oplus U_i\Sym^2V_i$.
These quotient operations remove the outputs of the two local-node
source spaces. The remaining central source $p\in Z_gZ_h$ projects
onto $U_iV_i$. Multiplication by $C_i$, whose $f$ coefficient is one,
then removes the four-dimensional $fU_iV_i$ summand.
This leaves
\begin{equation}\label{eq:doublecoker}
\mathcal C_i=\Sym^2U_i\otimes V_i\ \oplus\ U_i\otimes\Sym^2V_i,
\qquad \dim\mathcal C_i=12.
\end{equation}
Restricting the three-form Koszul syzygy \eqref{eq:koszul3} of
$(\ell_{g,i},\ell_{h,i},C_i)$ in Lemma~\ref{lem:koszul} to
$\tau\in Y_{g,h}$, $u\in Z_h$, and $v\in Z_g$ places
\begin{equation}\label{eq:doublekernel}
\begin{aligned}
q_g&=\ell_{h,i}\tau+C_i u,& q_h&=-\ell_{g,i}\tau+C_i v,\\
p&=-\ell_{g,i}u-\ell_{h,i}v
\end{aligned}
\end{equation}
inside the double-support source space, modulo the one-dimensional
relation $s(C_i,-\ell_{h,i},\ell_{g,i})$, which lies inside
$Y_{g,h}\times Z_h\times Z_g$ since $C_i\in Y_{g,h}$,
$\ell_{h,i}\in Z_h$, and $\ell_{g,i}\in Z_g$. Since $7+3+3-1=12$ matches
$\dim\mathcal C_i$, \eqref{eq:doublekernel} parametrizes the entire kernel.

Over an extension of $K$ (or adjoining indeterminates to $K$), choose
$\lambda,\mu\notin\{0,1\}$ with $\lambda\ne\mu$. Leave central $x$ fixed
and set
\begin{equation}\label{eq:doubleeta}
x_{g,i}=e_i+\eta(e_{i+1}+e_{i+2}),\qquad
x_{h,i}=e_i+\eta(\lambda e_{i+1}+\mu e_{i+2}).
\end{equation}
An edge from source $i$ to target $j=i+1$ uses weight $\theta=\lambda$,
and an edge to $j=i+2$ uses weight $\theta=\mu$.
Modulo target forms $\ell_{g,j},\ell_{h,j},C_j$, the first-order output is
\begin{equation}\label{eq:doubleedge}
(1-\theta)ab\,\tau+aC_i u+\theta bC_i v,
\end{equation}
where $a,b$ denote the quotient images of $\ell_{g,i},\ell_{h,i}$.
We must check both independence of the two outputs of one source and
complementarity of the two images at one target.

Both checks use the same three-point calculation. Put
$P_s=(c_{g,s},c_{h,s})$ and $\rho_s=(A_s-a_g)/(A_s-a_h)$.
The slope of $P_i-P_j$ is $\rho_i\rho_j$, by subtraction of reciprocals.
Since $a_g\ne a_h$, the fractional linear map $A_s\mapsto\rho_s$ is
injective and each $\rho_s$ is nonzero. Thus any three of the $P_s$
are affinely independent, in every characteristic.

\emph{Incoming edges.}
Fix target $j$ with incoming sources $i,k$. Take $a,a'$ along the
images of $\ell_{g,i},\ell_{g,k}$ in $U_j$ and $b,b'$ along those of
$\ell_{h,i},\ell_{h,k}$ in $V_j$. Normalize these basis vectors so that
$w_g=a+a'$ and $w_h=b+b'$. This is possible by the independence
properties following \eqref{eq:localforms}. In the quotient by $C_j$,
write $P_s-P_j=(d_s,e_s)$. Then
\[
C_s=d_s(a+a')+e_s(b+b'),\qquad s=i,k.
\]
By \eqref{eq:doubleedge}, the image from $i$ is
$T_{ij}=ab(U_j\oplus V_j)+aC_iV_j+bC_iU_j$; the image from $k$
has the same description with $a',b',C_k$.
The multiples of $ab$ and $a'b'$ account for eight distinct basis
monomials of $\mathcal C_j$. On the remaining pair $(a^2b',ab'^2)$,
the vectors $ab'C_i,ab'C_k$ have columns $(d_i,e_i),(d_k,e_k)$.
On the pair $(a'^2b,a'b^2)$, the vectors $a'bC_i,a'bC_k$ have
the same columns. Affine independence makes both pairs span.
Each edge contributes at most six dimensions (four multiples and two
extra vectors), so $\mathcal C_j=T_{ij}\oplus T_{kj}$ with both
summands of dimension six.

\emph{Outgoing edges.}
Fix source $i$ with targets $j=i+1$, $k=i+2$, and write
$N_s=\Span(\ell_{g,s},\ell_{h,s},C_s)$ for the space of linear
forms killed at target $s$. We claim that
\begin{equation}\label{eq:outgoingdecomposition}
Y_{g,h}=N_j\oplus N_k\oplus KC_i .
\end{equation}
Indeed, quotient first by the four local forms at $j,k$. The remaining
basis is $f,w_g,w_h$ by \eqref{eq:localforms}. In this basis
$C_j,C_k,C_i$ have coordinates $(1,P_j),(1,P_k),(1,P_i)$,
which are independent by the same three-point calculation.

Suppose $(\tau,u,v)$ gives zero output at both targets. At target $j$,
reduce \eqref{eq:doubleedge} modulo the image of $\ell_{h,i}$.
Only the term $\ell_{g,i}C_i u$ survives. Its first two factors are
nonzero in this polynomial ring (the $U_j$ component of $C_i$ is
nonzero), so $u\in\Span(\ell_{h,i},\ell_{h,j})$ in $Z_h$.
Target $k$ gives the same inclusion with $k$ in place of $j$.
Since the three local forms are independent, $u$ is a multiple of
$\ell_{h,i}$. Reducing instead modulo $\ell_{g,i}$ gives
$v\in K\ell_{g,i}$ in the same way.
Using the redundant line in \eqref{eq:doublekernel}, choose a
representative with $u=0$ and write $v=s\ell_{g,i}$.
Canceling the product of the two source forms in each target quotient
now gives
\[
 \tau+\frac{\lambda s}{1-\lambda}C_i\in N_j,\qquad
 \tau+\frac{\mu s}{1-\mu}C_i\in N_k .
\]
Subtracting and using \eqref{eq:outgoingdecomposition} gives $s=0$,
since the two scalar multipliers of $s$ differ. Then
$\tau\in N_j\cap N_k=0$. Thus the source map is injective, and the
dimension count $12=6+6$ makes it onto its two edge images.

Combining the source isomorphisms with the target direct sums proves that
the $48$-dimensional induced $\eta$ map is invertible.
Lemma~\ref{lem:lift} shows that the $180$-square determinant has a nonzero
coefficient of $\eta^{48}$.
Thus the double-support diagonal block is generically invertible.
\end{proof}

\subsection{Triple support: two four-dimensional tensor images}
\begin{lemma}[Triple-support nonvanishing]\label{lem:triple4}
Every $108$-square triple-support diagonal block has a determinant that
is a nonzero polynomial in its $x$ variables over the auxiliary parameter
field.
\end{lemma}
\begin{proof}
Fix three groups. There are no central sources in their $108$-square block.
At common $x_i=e_i$, each component is the $27$-square map
\[
(q_1,q_2,q_3)\longmapsto L_{1,i}q_1+L_{2,i}q_2+L_{3,i}q_3,
\]
with $q_1\in Z_2\otimes Z_3$ and cyclically.
By Lemma~\ref{lem:koszul}, the cokernel is
$\bigotimes_{g=1}^3(Z_g/\Span(L_{g,i}))$ of dimension $2^3=8$, and
restricting \eqref{eq:koszul3} to $\tau_{12}\in Z_3$, $\tau_{13}\in Z_2$,
$\tau_{23}\in Z_1$ gives the complete $(9-1=8)$-dimensional Koszul kernel
\begin{align}\label{eq:triplekernel}
q_1&=L_{2,i}\tau_{12}+L_{3,i}\tau_{13},\nonumber\\
q_2&=-L_{1,i}\tau_{12}+L_{3,i}\tau_{23},\\
q_3&=-L_{1,i}\tau_{13}-L_{2,i}\tau_{23},\nonumber
\end{align}
modulo $s(L_{3,i},-L_{2,i},L_{1,i})$.

Keep group one fixed and perturb groups two and three as the two groups in
\eqref{eq:doubleeta}. In the target quotient, the negative of the edge output is
\begin{equation}\label{eq:tripleedge}
\begin{aligned}
&L_{1,i}L_{2,i}\tau_{12}+\theta L_{1,i}L_{3,i}\tau_{13}\\
&\quad+(\theta-1)L_{2,i}L_{3,i}\tau_{23},\qquad \theta=\lambda,\mu.
\end{aligned}
\end{equation}
For source $i$ with targets $j,k$, use bases $(L_{g,i},L_{g,j},L_{g,k})$.
The output at $j$ kills the $L_{g,j}$ coefficients and separately detects
the three $L_{g,k}$ coefficients; the output at $k$ detects the three
$L_{g,j}$ coefficients. Their multipliers are $1,\theta,\theta-1$,
all nonzero. The three remaining $L_{g,i}$ coefficients are read by
the two independent rows
\[
(1,\lambda,\lambda-1),\qquad(1,\mu,\mu-1).
\]
Their common kernel is exactly the redundant line $(1,-1,1)$.
The difference $(0,\lambda-\mu,\lambda-\mu)$ proves the rows independent
in every characteristic.
Thus the source kernel maps isomorphically onto its two four-dimensional
edge images.

At a target $j$ with incoming sources $i,k$, the quotient bases in each
group are $(L_{g,i},L_{g,k})$. The image of source $i$ is spanned by the
four tensor monomials with at least two $i$ factors. The image of source
$k$ has at least two $k$ factors. These are disjoint sets comprising
all eight basis tensors. The $32$-dimensional induced map is therefore
invertible, and the $108$-square determinant has nonzero $\eta^{32}$ coefficient.
This supplies the last larger-support block needed for elimination.
We now return to the actual Vandermonde parameter $\eps$.
\end{proof}

\subsection{Assembly for arbitrary $r$ and the global first-order lift}\label{subsec:assembly}
Figure~\ref{fig:globallift} shows how the support elimination and the
cokernel projection fit together. Lemma~\ref{lem:supportelimination} gives the kernel corrections;
Proposition~\ref{prop:globalassembly} then identifies the entire
first-order map on the missing conditions.

\begin{proofmapfigure}
\centering
\setlength{\fboxsep}{3pt}
\fbox{\parbox[c][46pt][c]{0.18\textwidth}{\centering\small
Singleton kernel\\$\ker B_{\{g\}}$\\dimension $4$}}
$\longrightarrow$
\fbox{\parbox[c][46pt][c]{0.18\textwidth}{\centering\small
Unique corrections\\$q_T$, $|T|=2,3$\\only if $g\in T$}}
$\longrightarrow$
\fbox{\parbox[c][46pt][c]{0.18\textwidth}{\centering\small
Cokernel test\\$\Phi_{h,i}(\A_1q)$\\reads pure $h$}}
$\longrightarrow$
\fbox{\parbox[c][46pt][c]{0.18\textwidth}{\centering\small
Groupwise lift\\$\bigoplus_g\mathcal L_g$\\invertible $4$-blocks}}
\caption{The global $t=4$ lift. A kernel vector from group $g$
acquires corrections on larger supports containing $g$. By
\eqref{eq:puresupportformula}, the test at $h$ reads only a central
quadratic supported on $\{h\}$. Thus it gives zero for $h\ne g$
and the one-group lift for $h=g$.}
\label{fig:globallift}
\end{proofmapfigure}

Let $\pi_g:Y\to V_g=\Span(z_{g,0},z_{g,1})$ kill $f$, all $w$ vectors,
and the other groups. Let $\xi_{g,i}$ be dual to the local basis $x_{g,i}$,
and put $\ell_i(z_{g,0})=-b_i$, $\ell_i(z_{g,1})=1$.
Define the pure-support functional
\begin{equation}\label{eq:globalcoker}
\Phi_{g,i}(x\otimes uvw)
=\xi_{g,i}(x)\ell_i(\pi_gu)\ell_i(\pi_gv)\ell_i(\pi_gw).
\end{equation}
The expression is symmetric, so it defines a functional on the ordinary
symmetric power in every characteristic. It annihilates $\A_0$:
central and other-group node forms project to zero, and the surviving
local form $z_{g,0}+b_i z_{g,1}$ is killed by $\ell_i$.
Also $\Phi_{g,i}(x_{g,j}\otimes z_{g,1}^3)=\delta_{ij}$, which shows
independence within a group; different groups read disjoint supports.

\begin{lemma}[Pure-support reduction of the first-order map]\label{lem:puresupportlift}
For the family \eqref{eq:jets} and $q\in\ker\A_0$, the full first-order
map satisfies
\begin{equation}\label{eq:puresupportformula}
\Phi_{g,i}(\A_1q)=\sum_{j=0}^3
\xi_{g,i}(x_{0,j})c_{g,j}^2
\ell_i^{(2)}\bigl(\pi_g^{(2)}q_{0,j}\bigr).
\end{equation}
Here $q_{0,j}$ is the quadratic source at central helper $j$,
$\pi_g^{(2)}$ is the induced projection on quadratics, and
$\ell_i^{(2)}(uv)=\ell_i(u)\ell_i(v)$.
Under the hypotheses of Lemma~\ref{lem:supportelimination}, a kernel
vector extended from singleton $h$ is invisible in \eqref{eq:puresupportformula}
when $g\ne h$; when $g=h$, its value is the one-group first-order map.
\end{lemma}
\begin{proof}
Project $\A_0q=0$ to $X\otimes\Sym^3 V_g$. Only local group $g$ remains:
\[
\sum_{i=0}^3x_{g,i}\otimes(z_{g,0}+b_i z_{g,1})\pi_g^{(2)}q_{g,i}=0.
\]
The $x_{g,i}$ are a basis, and multiplication by each displayed
nonzero linear form is injective. Thus
\begin{equation}\label{eq:projectedkernelzero}
\pi_g^{(2)}q_{g,i}=0\qquad(0\leq i<4).
\end{equation}

To track the coordinate-change term, let $D_g:Y\to Y$ send $w_g$ to
$z_{g,1}$ and every other basis vector to zero. Extend it by the
Leibniz rule to $D_g^{(j)}$ on $\Sym^jY$. Changing each node form by
$D_g y$ induces the matrix variation
\begin{equation}\label{eq:coordinatevariation}
\dot\A_g=(I_X\otimes D_g^{(3)})\A_0
-\A_0\Bigl(\bigoplus_hD_g^{(2)}\Bigr).
\end{equation}
Indeed, $D_g^{(3)}(yq)=(D_gy)q+yD_g^{(2)}q$ for each quadratic $q$.
Both terms in \eqref{eq:coordinatevariation} vanish after applying
$\Phi_{h,i}$ to a vector in $\ker\A_0$.
The node changes comprising $\dot\A_g$ are central
$c_{g,j}z_{g,1}$ and local $b_i^2z_{g,1}$.
The local contribution is already zero by \eqref{eq:projectedkernelzero}
for $h=g$, and projects to zero for $h\ne g$. Hence the central
contribution alone has zero induced map, as does its multiple $-S_g$.
This explicitly removes the central $-S_gc_{g,j}z_{g,1}$ terms.

All actual local first-order motions in \eqref{eq:jets} lie in the span
of $f$ and the $w$ vectors, so every $\Phi_{g,i}$ kills them directly.
The remaining central motion is $c_{g,j}^2z_{g,1}$; since
$\ell_i(z_{g,1})=1$, its evaluation is exactly
\eqref{eq:puresupportformula}.

Finally, the support recursion extends a singleton-$h$ kernel only
through source labels containing $h$. A central quadratic surviving
$\pi_g^{(2)}$ has label $\{g\}$. Thus neither an $h\ne g$ source
nor any higher-support correction enters the formula. For $h=g$,
the central pure-$g$ source is unchanged by those corrections and
the formula coincides with the one-group map of Lemma~\ref{lem:single4}.
\end{proof}

\begin{proposition}[Global support assembly]\label{prop:globalassembly}
Consider the normalized family $\A(\eps)=\A_0+\eps\A_1+O(\eps^2)$
defined by \eqref{eq:jets}, with $G=r-1$ local groups. Choose generic
first factors satisfying Lemmas~\ref{lem:single4}--\ref{lem:triple4},
with the central and each local four-tuple independent. Then
$\ker\A_0$ is the direct sum of the $G$ four-dimensional singleton
kernels, extended uniquely through the higher-support equations.
The pure-support functionals in \eqref{eq:globalcoker} form a basis
of $(\coker\A_0)^*$, and in these coordinates the induced first-order map
is the direct sum of the invertible one-group maps $\mathcal L_g$.
Consequently $\A_0$ has corank $4G$ and $\det\A(\eps)$ has order $4G$.
\end{proposition}
\begin{proof}
Work over the common rational function field in the node parameters
and first factors. Each local witness certifies a
nonzero minor or determinant coefficient in the same parameter ring,
after clearing node denominators. Their product, including the basis
determinants of the central and local four-tuples, is nonzero.

Apply Lemma~\ref{lem:supportelimination} to Table~\ref{tab:blocks}.
Each singleton kernel contributes four free coordinates, so
\begin{equation}\label{eq:globalcorank}
\dim\ker\A_0=\dim\coker\A_0=4G.
\end{equation}
The $4G$ independent functionals \eqref{eq:globalcoker} therefore
give coordinates on the cokernel.

Consider first the running case $r=3$. The two singleton inputs
$q_1,q_2$ determine one correction by \eqref{eq:supportrecursion},
\[
B_{\{1,2\}}q_{\{1,2\}}=-\sum_{g=1}^2 A_{\{1,2\},\{g\}}q_g .
\]
By \eqref{eq:puresupportformula}, the four tests at group $1$ read
only central quadratics supported on $\{1\}$. Thus they see $q_1$
alone, even after the correction has been added. The four tests at
group $2$ likewise see only $q_2$. The induced map on the eight
missing conditions is consequently
\[
\begin{pmatrix}\mathcal L_1&0\\0&\mathcal L_2\end{pmatrix},
\]
with two four-dimensional diagonal blocks. Appendix~\ref{app:liftcertificate}
evaluates this matrix over $\F_{101}$ for the running example.

For general $r$, extend an input from singleton $h$. By
Lemma~\ref{lem:supportelimination}, every nonzero component of this
extension has a support $T$ containing $h$. To contribute to the test
at $g$, its central quadratic must also satisfy $T=\{g\}$ by
\eqref{eq:puresupportformula}. Together these conditions force
$g=h$ and $T=\{h\}$. This excludes every pair correction and,
starting at $r=4$, every triple correction. The surviving block is
the one-group lift of Lemma~\ref{lem:single4}. Hence
\begin{equation}\label{eq:globallift}
\overline{\A}_1=\bigoplus_{g=1}^G\mathcal L_g .
\end{equation}
All summands are invertible, so Lemma~\ref{lem:lift} gives order
$4G$ for the normalized determinant.
\end{proof}

\begin{proof}[Proof of Theorem~\ref{thm:t4}]
Proposition~\ref{prop:globalassembly} certifies a nonzero determinant for
the normalized family over the rational function extension $K$ of $F$.
Because the Hermite coordinate change and node scalings are invertible
over $K((\eps))$, the original Vandermonde repair determinant is nonzero
under this specialization and hence is a nonzero polynomial over $F$.
Homogeneity of degree $\beta$ in each $x_h$ preserves nonvanishing after
setting $x_{h,0}=1$.
\end{proof}

\subsection{Higher-Degree Paired-Channel Cases at $r=2$: $t=5,6$}\label{subsec:higherr2}
At $r=2$ ($d=2t$, $k=2t-1$, $m=t$, $\alpha=\binom{2t-2}{t-1}$,
$\beta=\alpha/2$), grouping the $2t$ fixed MDS second-factor vectors
into $t$ pairs gives two-form Koszul kernels
$\{(L_i\ell_i,-F_i\ell_i):\ell_i\in\Sym^{t-3}Y\}$ without higher
syzygies in the baseline blocks. The same paired-channel reduction
extends fixed-MDS nonvanishing to $t=5$ and $t=6$.

\begin{theorem}[Fixed MDS nodes for $t=5,6$, $r=2$]\label{thm:higherrepair}
Let $t\in\{5,6\}$ and let $X,Y$ be $t$-dimensional over any field $F$.
Fix $2t$ vectors $y_h\in Y$ such that every $t$ are independent.
Then the determinant of
\[
\begin{aligned}
\A&:\bigoplus_{h=1}^{2t}\Sym^{t-2}Y\longrightarrow X\otimes\Sym^{t-1}Y,\\
&(q_h)\longmapsto\sum_{h=1}^{2t}x_h\otimes y_hq_h
\end{aligned}
\]
is a nonzero polynomial in the free coordinates of the $x_h$ over $F$.
\end{theorem}
\noindent The proofs are given in Appendices~\ref{app:t5} and~\ref{app:t6}.

\begin{corollary}[One code for every helper set at $t=5,6$, $r=2$]\label{cor:highercode}
For $t\in\{5,6\}$, $n\geq2t+1$, and any finite field of order
\begin{equation}\label{eq:higherD}
q>D_t(n):=\binom{n-1}{t-1}+\beta\binom{n-1}{2t-1},
\end{equation}
there is a linear all-node, arbitrary-helper exact-repair MSR code with
\[
\begin{array}{c|rrrr}
t&k&d&\alpha&\beta\\\hline
5&9&10&70&35\\
6&11&12&252&126
\end{array}
\]
One node configuration supports every reconstruction and repair condition.
\end{corollary}
\begin{proof}
The bound ensures $q\geq n$, since
$D_t(n)\geq\binom{n-1}{t-1}\geq n-1$ for $n\geq2t+1$.
Fix distinct $a_i$ and $y_i=(1,a_i,\ldots,a_i^{t-1})$, so every $t$
are independent. Normalize $x_i=(1,u_{i,1},\ldots,u_{i,t-1})$ and form
\begin{equation}\label{eq:higherproduct}
P_t=\prod_{|I|=t}\det[x_i]_{i\in I}\ \prod_{|H|=2t}\det\A_H.
\end{equation}
Every factor is nonzero, the repair factors by
Theorem~\ref{thm:higherrepair}, and normalizing the leading coordinate
preserves nonvanishing by homogeneity in each node. Each free coordinate
of $x_i$ occurs in $\binom{n-1}{t-1}$ independence minors at degree at
most one and in $\binom{n-1}{2t-1}$ repair determinants at degree at
most $\beta$, so \eqref{eq:higherD} bounds its individual degree.
Lemma~\ref{lem:grid} gives a common point with $P_t\ne0$, and
Proposition~\ref{prop:code} yields reconstruction and exact repair.
\end{proof}

At shortest lengths, $D_5(11)=560$ and $D_6(13)=2304$, so $\F_{1024}$
and $\F_{4096}$, respectively, suffice.

\section{Small-Field Selection and Explicit Extension-Field Coordinates}\label{sec:fields}
The nonvanishing theorems must now yield one configuration satisfying
every reconstruction and helper-set condition. We multiply the
corresponding nonzero determinants and apply the standard finite-grid
lemma below. The individual-degree counts give explicit sufficient
field orders. Theorem~\ref{thm:detselection} gives the cost of
deterministic selection.

\begin{lemma}[Finite-grid nonvanishing]\label{lem:grid}
Let $P\neq0$ be a polynomial over $F$ with degree at most $D_j$ in variable
$z_j$. If $S_j\subset F$ has more than $D_j$ elements, $P$ is nonzero at
some point of $\prod_jS_j$.
\end{lemma}
\begin{proof}
Induct on the number of variables. The one-variable case is the root bound.
For the induction step, write
$P=\sum_{e=0}^{D_1}z_1^eP_e(z_2,\ldots,z_N)$ and choose a nonzero
coefficient $P_e$. It satisfies the same individual-degree bounds for
the remaining variables. By induction, those variables can be assigned
grid values for which $P_e$ is nonzero. The residual polynomial in $z_1$
is then nonzero and has degree at most $D_1$, so it cannot vanish on all
of $S_1$. Choosing a nonroot finishes the induction.
\end{proof}

\begin{theorem}[Finite fields for $t=3$]\label{thm:field3}
For $r\geq2$ and $n\geq3r+1$, put
\begin{equation}\label{eq:D3}
D_3=\binom{n-1}{2}+(2r-1)\binom{n-1}{3r-1}.
\end{equation}
Over every finite field $F$ of order $q>D_3$, fix any $n$ vectors
$y_i\in F^m$ such that every $m$ are independent. In particular, one may
choose distinct $a_i$ and take
\[
y_i=(1,a_i,\ldots,a_i^{2r-2}).
\]
There exist $x_i=(1,u_i,v_i)$ satisfying all conditions of
Proposition~\ref{prop:code}; they can be selected by a deterministic finite procedure.
\end{theorem}
\begin{proof}
The field has enough elements: for $n\geq7$, $D_3\geq\binom{n-1}{2}\geq n$.
Thus the Vandermonde choice is available. For any fixed family satisfying
the stated independence condition, let
\[
P_3=\prod_{|I|=3}\det[x_i]_{i\in I}\ \prod_{|H|=3r}\det\A_H.
\]
Every factor is a nonzero polynomial by Theorem~\ref{thm:t3} or by the
elementary determinant for three free vectors, so their product in the
common polynomial ring is nonzero. At any point where $P_3\ne0$, every
triple of $x$ vectors and every helper-set determinant is nonsingular.
Each scalar $u_i$ or $v_i$ occurs in $\binom{n-1}{2}$ vector determinants
at degree at most one and $\binom{n-1}{3r-1}$ repair determinants at
degree at most $\beta=2r-1$, giving \eqref{eq:D3}. Lemma~\ref{lem:grid}
and Proposition~\ref{prop:code} complete the proof.
\end{proof}

\begin{theorem}[Finite fields for $t=4$]\label{thm:field4}
For $r\geq2$ and $n\geq4r+1$, put $\beta=\binom{3r-1}{2}$ and
\begin{align}\label{eq:D4}
D_x&=\binom{n-1}{3}+\beta\binom{n-1}{4r-1},\\
D_a&=n-1+(3r-3)\beta\binom{n-1}{4r-1}.\nonumber
\end{align}
Every finite field with $q>\max(D_x,D_a)$ admits pairwise distinct $a_i$
and $x_i=(1,u_i,v_i,w_i)$ satisfying Proposition~\ref{prop:code}.
Moreover, at $(t,r)=(4,2)$, for any prescribed $n$ vectors $y_i\in F^4$
with every four independent, every finite field with $q>D_x$ admits
$x_i=(1,u_i,v_i,w_i)$ satisfying Proposition~\ref{prop:code}. Both
selections are deterministic and finite.
\end{theorem}
\begin{proof}
For joint Vandermonde selection, form the nonzero product
\[
P_4=\prod_{i<j}(a_i-a_j)\ \prod_{|I|=4}\det[x_i]_{i\in I}
\ \prod_{|H|=4r}\det\A_H,
\]
using Theorem~\ref{thm:t4}. Each node supplies $\beta$ repair columns of
degree at most $m-1=3r-3$ in $a_i$ and degree at most one in $x_i$, giving
the individual-degree bounds \eqref{eq:D4}; Lemma~\ref{lem:grid} applies
jointly to all $4n$ free scalar coordinates. At $(t,r)=(4,2)$ with fixed
MDS $y_i$, Proposition~\ref{prop:t4r2} makes
$\prod_{|I|=4}\det[x_i]_{i\in I}\prod_{|H|=8}\det\A_H$ a nonzero
polynomial of individual degree at most $D_x$ in the $3n$ free
coordinates of $x_i$.
\end{proof}

\begin{theorem}[Deterministic node selection]\label{thm:detselection}
Given the finite field in an arithmetic representation with an enumeration
of its elements, the coordinates in Theorems~\ref{thm:field3}
and~\ref{thm:field4} can be constructed deterministically at their stated
field sizes using polynomially many field operations in $n$ and $\binom nd$.
In particular, selection takes polynomially many field operations in $n$
at fixed $c=n-d$, including $c=1$.
\end{theorem}
The algorithm starts with a separate nonzero evaluation for each
constraint. It fixes one scalar at a time, preserving all those
evaluations on the unfixed coordinates. The constructive step is to
obtain the initial evaluations from the two determinant proofs;
Appendix~\ref{app:selector} gives the algorithm and its complexity.

\begin{corollary}[Shortest length and fixed excess length]\label{cor:asymptotic}
At $n=d+1$, sufficient thresholds are
\begin{align}\label{eq:shortfields}
D_3&=\frac{3r(7r-3)}2,\\
\max(D_x,D_a)&=D_a=4r\left[1+(3r-3)\binom{3r-1}{2}\right].\nonumber
\end{align}
For any prescribed prime characteristic $p$, the smallest power of $p$
strictly greater than a threshold $D$ is at most $pD$.
Thus $q=O_p(n^2)$ for $t=3$ and $q=O_p(n^4)$ for $t=4$.
For fixed $n-d=c\geq1$, the bounds are
$O_{p,c}(n^{c+1})$ and $O_{p,c}(n^{c+3})$.
Choosing characteristic two gives the usual multiplier of at most two.
\end{corollary}
\begin{proof}
At shortest length the repair-subset count is $\binom{tr}{tr-1}=tr$.
Substitution gives \eqref{eq:shortfields}; for $r\geq2$ the displayed
$D_a$ exceeds $D_x$. At fixed excess $c$, the count is
$\binom{n-1}{d-1}=\binom{n-1}{c}=O_c(n^c)$, while
$\beta=O(r)$ for $t=3$ and $O(r^2)$ for $t=4$.
\end{proof}
If $n-d$ varies with $n$, the helper-subset counts in the field bounds
remain explicit. The case $r=1$ is elementary: $k=d=t$ and
$\alpha=\beta=1$, so a scalar MDS code suffices.

\begin{example}[Two shortest-length codes over nine elements]\label{ex:nine}
Write $\F_9=\F_3[u]/(u^2+1)$ and encode $a+bu$ by the integer $a+3b$,
with $a,b\in\{0,1,2\}$. At $(t,r,n)=(3,2,7)$ let $a_i$ have labels
$0,\ldots,6$, take $y_i=(1,a_i,a_i^2)$, and use $x_i=(1,v_i,w_i)$ with
\[
(v_i)=(6,7,2,4,6,5,1),\quad (w_i)=(4,3,8,3,2,4,7)
\]
in the same labeling. At $(t,r,n)=(4,2,9)$ index nodes by
$z\in\F_9$. With $\eta=1+u$ use the projection of Glynn's
arc~\cite{glynn1986} for the first factors,
\[
x_z=(1,z,z^2+\eta z^6,z^3),\qquad
y_z=(1,a_z,a_z^2,a_z^3),
\]
where $a_z$ interchanges $0,1$ and fixes the other seven elements.
Direct evaluation over $\F_9$ checks all $35$ and $126$ first-factor
minors and all $7$ and $9$ repair determinants, respectively, while
distinct $a_z$ guarantee all second-factor minors. Proposition~\ref{prop:code} yields both codes, and
for $(4,2,9)$ the field size $q=9$ attains the trivial bound $q\geq n$ for
distinct affine Vandermonde parameters.

For $(3,2,7)$, $q=9$ is likewise minimal under affine Vandermonde second
factors and Proposition~\ref{prop:code}. Distinct $a_i$ require $q\geq7$,
leaving $q\in\{7,8\}$, while three-wise independent first factors
$x_1,\ldots,x_7\in\F_q^3$ define a $7$-arc in $\mathrm{PG}(2,q)$. In
$\mathrm{PG}(2,7)$, every $7$-arc extends to an $8$-arc, which is a
nonsingular conic by Segre's theorem~\cite[Thm.~8.14 and
Sec.~14.5]{hirschfeld1998}; in $\mathrm{PG}(2,8)$, every $7$-arc extends
to a $10$-point hyperoval, and every hyperoval in $\mathrm{PG}(2,8)$ is
regular (a $9$-point nonsingular conic together with its nucleus), so at
least six of the seven points lie on a nonsingular
conic~\cite[Sec.~14.6]{hirschfeld1998}. Over
$\overline{\F}_q$, an invertible linear change on $X$ and nonzero node
scalings map those six first factors to $(1,z_h,z_h^2)$ with finite $z_h$
while preserving repair rank, so Theorem~\ref{thm:curveobstruction}
forces their six-helper repair determinant to vanish.
\end{example}

\subsection{Randomized selection at fixed excess length}\label{subsec:randomselection}
For a finite field given with an arithmetic representation, the same
constraint products yield a Las Vegas algorithm: sample the free node
coordinates and accept only after every required determinant is nonzero.
The total-degree bound controls the number of trials, whereas the
individual-degree bounds above give the smaller thresholds for deterministic selection.

\begin{corollary}[Las Vegas node selection]\label{cor:randomselection}
Let $t\in\{3,4\}$, $r\geq2$, $n\geq d+1$, and $d=tr$. Put
\begin{align}\label{eq:randomdegrees}
\Delta_3&=2\binom n3+d\beta\binom nd,\\
\Delta_4&=\binom n2+3\binom n4+md\beta\binom nd.\nonumber
\end{align}
Over any finite field of order $q>2\Delta_t$, independent uniform
sampling of the free coordinates, followed by exact verification,
outputs a configuration satisfying Proposition~\ref{prop:code} with
success probability greater than $1/2$ per trial and expected fewer
than two trials. For $t=3$, any $n$ distinct Vandermonde parameters
are fixed before sampling the first factors; for $t=4$, the parameters
and first factors are sampled jointly.

For fixed $c=n-d\geq1$, the expected number of field operations is
polynomial in $n$. In each prescribed characteristic, sufficient field
orders are $O_{p,c}(n^{c+2})$ for $t=3$ and
$O_{p,c}(n^{c+4})$ for $t=4$; at $c=1$ these become $O_p(n^3)$ and
$O_p(n^5)$.
\end{corollary}
\begin{proof}
The products $P_3,P_4$ in Theorems~\ref{thm:field3}--\ref{thm:field4}
are nonzero. A determinant of $t$ rows $(1,x_{i,1},\ldots,x_{i,t-1})$
has total degree at most $t-1$. Each repair determinant has $d\beta$
columns. For $t=3$ each entry has degree at most one in the sampled
coordinates; for $t=4$ an entry $x_{h,i}a_h^b$ has joint degree at
most $m$. Adding the $\binom n2$ parameter differences for $P_4$
gives \eqref{eq:randomdegrees}.

A nonzero polynomial of total degree at most $\Delta$ vanishes at
independent uniform field elements with probability at most $\Delta/q$.
For completeness, write it as a polynomial of degree $e$ in its last
variable. The leading coefficient has total degree at most
$\Delta-e$; induction bounds the probability that this coefficient
vanishes by $(\Delta-e)/q$, and the univariate root bound contributes
at most $e/q$ on the remaining assignments. This proves the claim.
Applied to $P_t$, it gives rejection probability less than $1/2$.
One need not expand the product: test its factors separately by exact
linear algebra and reject at the first zero. Every accepted sample
therefore gives a valid code, and independent repetition has expected
fewer than two trials.

There are $\binom nt$ first-factor minors, $\binom nd$ repair
determinants of order $d\beta=t\alpha$, and, for $t=4$, pairwise
parameter-distinctness tests. Matrix formation and Gaussian elimination
give a polynomial field-operation cost in $n$ and $\binom nd$ for
fixed $t$. When $c$ is fixed, $\binom nd=\binom nc=O_c(n^c)$ and
$d\beta=O(n^{t-1})$. Equation~\eqref{eq:randomdegrees} then gives
$\Delta_3=O_c(n^{c+2})$ and $\Delta_4=O_c(n^{c+4})$.
The least power of a prescribed prime $p$ exceeding $2\Delta_t$ is
at most $2p\Delta_t$, proving the field-order statements.
\end{proof}

This algorithm uses the field representation as input. In characteristic
two, the explicit tower in \eqref{eq:binarytower} also supplies one:
the least tower order $q=2^{2^b}>2\Delta_t$ satisfies
$q\leq(2\Delta_t)^2$. Thus fixed excess length also admits a fully
specified binary field of polynomial order for randomized selection.

\paragraph{Verification complexity}
With $N=d\beta=t\alpha$, dense elimination gives the per-trial bound
\begin{equation}\label{eq:verificationcost}
O\!\left(\binom ndN^3+\binom nt t^3+n^2\right)
\end{equation}
field operations, which also bounds expected selection cost since fewer
than two trials are needed in expectation. For fixed $c=n-d$ this gives
\[
t=3:\ O_c(n^{c+6}),\qquad t=4:\ O_c(n^{c+9}),
\]
specializing at $c=1$ to $O(n^7)$ and $O(n^{10})$.

\subsection{Explicit binary coordinates via helper-set interpolation}\label{subsec:interpolatednodes}
Whereas Theorems~\ref{thm:field3}--\ref{thm:detselection} search for
coordinates over small fields, Kronecker degree separation yields closed-form
coordinates over larger specified extension fields without candidate
testing. Interpolating the local perturbation pencils across the
$J=\binom nd$ helper sets keeps the extension degree over $\F_2$
polynomial in $n$ whenever $c=n-d$ is fixed.

\begin{lemma}[Degree-separated evaluation]\label{lem:kronecker}
Let $\mathcal P$ be a family of nonzero polynomials in
$K[w_1,\ldots,w_N]$, each satisfying $\deg_{w_j}P\leq d_j$.
Set $W_j=\prod_{\ell<j}(d_\ell+1)$ and
$L=\prod_j(d_j+1)$. If $[K(\theta):K]\geq L$, then
$P(\theta^{W_1},\ldots,\theta^{W_N})\ne0$ for every $P\in\mathcal P$.
\end{lemma}
\begin{proof}
For each $P$, distinct monomials map to distinct exponents
$\sum_j e_jW_j\in\{0,\ldots,L-1\}$ by uniqueness of mixed-radix
representation, so the substituted univariate polynomial is nonzero of
degree less than $L$ and cannot vanish at $\theta$.
\end{proof}

To fix an explicit field representation in characteristic two, start with
$K_0=\F_2$, $\delta_0=1$, and recursively set
\begin{equation}\label{eq:binarytower}
\begin{split}
K_\ell&=K_{\ell-1}[T]/(T^2+T+\delta_{\ell-1}),\\
\theta_\ell&=T\bmod(T^2+T+\delta_{\ell-1}),\qquad
\delta_\ell=\delta_{\ell-1}\theta_\ell.
\end{split}
\end{equation}
Inductively $\operatorname{Tr}_{K_{\ell-1}/\F_2}(\delta_{\ell-1})=1$
(starting from $\operatorname{Tr}_{K_0/\F_2}(\delta_0)=1$), so
$T^2+T+\delta_{\ell-1}$ is irreducible over $K_{\ell-1}$; its root
$\theta_\ell$ satisfies $\operatorname{Tr}_{K_\ell/K_{\ell-1}}(\theta_\ell)=1$,
whence $\operatorname{Tr}_{K_\ell/K_{\ell-1}}(\delta_\ell)=\delta_{\ell-1}$
and $\operatorname{Tr}_{K_\ell/\F_2}(\delta_\ell)=1$. Thus
$|K_\ell|=2^{2^\ell}$, and because $\theta_\ell\notin K_{\ell-1}$ (the
unique maximal subfield of $K_\ell$), $\theta_\ell$ generates $K_\ell$
over $\F_2$. Enumerate the basis of $K_b$ by
$e_j=\prod_{\ell=1}^b\theta_\ell^{\operatorname{bit}_{\ell-1}(j)}$
($0\leq j<2^b$) and assign index $h$ the field element
$\gamma_h=\sum_{j<2^b}\operatorname{bit}_j(h)e_j$.

\begin{theorem}[Interpolated binary coordinates]\label{thm:interpolatednodes}
Let $t,r,n$ satisfy Theorem~\ref{thm:main}, and put $N=d\beta$ and
$J=\binom nd$. Let $K_b$ be the first field in \eqref{eq:binarytower}
with $|K_b|\geq\max(n,J+1)$. The formulas below give an MSR code over
$K_{b+\lceil\log_2 L\rceil}$, where
\begin{align*}
L&=(NJ+1)(N+1) &&(t=3),\\
L&=(mNJ+1)(2N+1)^2(N+1)\\
 &\qquad{}\cdot((m-1)N+1) &&(t=4).
\end{align*}
For fixed $n-d$, the absolute extension degree over $\F_2$ is polynomial
in $n$ (field order $2^{\mathrm{poly}(n)}$).
\end{theorem}

\emph{Node coordinates.}
With $\gamma_h\in K_b$ ($0\leq h<n$) as above and
$y(a)=(1,a,\ldots,a^{m-1})$, index the $J+1$ interpolation points by
$\star$ followed by all $d$-subsets in lexicographic order, assign them
$\tau_l=\gamma_l$ ($0\leq l\leq J$), and set
\[
\Lambda_l(U)=\prod_{k\ne l}(U-\tau_k)/(\tau_l-\tau_k).
\]
At $\star$ prescribe $X_{h,\star}=(1,\gamma_h,\ldots,\gamma_h^{t-1})$
and $a_{h,\star}=\gamma_h$. At label $H$, retain these prescriptions
for $h\notin H$; for $h\in H$ list the helpers in increasing order and set:
\begin{itemize}
\item For $t=3$, split the list into three successive groups of $r$.
For position $j=0,\ldots,r-1$ in group $g=0,1,2$, put
\[
X_{h,H}=e_g+S\gamma_j e_{g+1},\qquad a_{h,H}=\gamma_h,
\]
with basis indices modulo three.
\item For $t=4$, split the list into $r$ successive groups of four,
numbered $g=0,\ldots,r-1$ with group zero central. For position
$i=0,\ldots,3$, put $p_g=V+W\gamma_g$ and set
\begin{align*}
X_{h,H}&=e_i+S(p_ge_{i+1}+p_g^2e_{i+2}),\\
a_{h,H}&=
\begin{cases}
\gamma_i,&g=0,\\
\gamma_{3+g}+E\gamma_i,&g>0,
\end{cases}
\end{align*}
with basis indices modulo four.
\end{itemize}
Define
\begin{equation}\label{eq:interpolatedcoords}
(x_h,a_h)=\sum_l\Lambda_l(U)(X_{h,l},a_{h,l}),\qquad y_h=y(a_h),
\end{equation}
and substitute powers $\theta^{W_j}$ of
$\theta=\theta_{b+\lceil\log_2 L\rceil}$ as in Lemma~\ref{lem:kronecker}
using ordered variables $(U,S)$ with bounds $(NJ,N)$ for $t=3$, and
$(U,V,W,S,E)$ with bounds $(mNJ,2N,2N,N,(m-1)N)$ for $t=4$.

\begin{proof}
At $U=\tau_\star$, every first-factor minor and parameter difference is
nonzero. At $U=\tau_H$, the helpers in $H$ specialize for $t=3$ to the
pencil \eqref{eq:t3pencil} with distinct weights $\gamma_j$, so
$\det\A_H$ is nonzero in $S$.

For $t=4$, treat $V,W,S,E$ as independent indeterminates. At $E=0$, each
group's four first factors equal the standard basis at $S=0$ and so are
generically independent. Assigning $(V,W)$ so that $\gamma_g\mapsto p_g$
maps $\{0,g\}$ to $\{1,0\}$ recovers \eqref{eq:singleeta} with $\eta=S$
for any singleton $g$. Mapping $\{0,g\}$ to $\{0,1\}$ sends any second
local group $h\ne g$ to $\lambda\in K_b\setminus\{0,1\}$; in
characteristic two, $\lambda$ and $\lambda^2$ are distinct and lie
outside $\{0,1\}$, recovering \eqref{eq:doubleeta}. Mapping the first two
groups of a triple to $\{0,1\}$ similarly gives weights
$\lambda,\lambda^2$ on the third group for Lemma~\ref{lem:triple4}.
Proposition~\ref{prop:globalassembly} and the invertible Hermite change
therefore make $\det\A_H$ a nonzero polynomial in $V,W,S,E$ at $U=\tau_H$.

Each repair determinant has $N$ columns. For $t=3$, $a_h=\gamma_h$ is
constant in $U$, giving entry degrees at most $(J,1)$ in $(U,S)$. For
$t=4$, each entry $x_{h,i}a_h^j$ ($j<m$) has degrees at most
$(mJ,2,2,1,m-1)$ in $(U,V,W,S,E)$. Multiplying by $N$ covers all repair,
reconstruction, and parameter-difference factors, so
Lemma~\ref{lem:kronecker} and Proposition~\ref{prop:code} establish the
code over $K_{b+\lceil\log_2 L\rceil}$. Finally,
$[K_b:\F_2]<2\max(n,J+1)$ and the relative extension degree is less than
$2L$, which is polynomial in $n$ for fixed $c=n-d$.
\end{proof}

\begin{remark}[Direct Kronecker coordinates and small-field specializations]\label{rem:kroneckergeneral}
When $n-d$ grows arbitrarily, applying Lemma~\ref{lem:kronecker} directly
to the $2n$ (for $t=3$ over $K_b$ with $|K_b|\geq n$) or $4n$ (for $t=4$
over $\F_2$) free coordinates of $P_t$ avoids enumerating the $J$ helper
sets, using extension thresholds $L_3=(2r)^{2n}$ and
$L_4=((3r-3)\beta+1)^n(\beta+1)^{3n}$ at the cost of doubly exponential
field order. Conversely, for concrete small parameters, evaluating
\eqref{eq:interpolatedcoords} at a single tested element $\theta\in K_3=\F_{256}$
yields compact explicit codes: in the basis enumeration above, $\theta$
with index $18,18,16,75$ realizes
$(t,r,n)=(3,2,7),(4,2,9),(3,3,10),(4,3,13)$ over $\F_{256}$, respectively.
\end{remark}

\section{An Obstruction to Consecutive-Power First-Factor Curves}
\label{sec:obstruction}
Consider the specific first-factor curve
$x_h=(1,z_h,\ldots,z_h^{t-1})$, together with Vandermonde second
factors $y_h=y(a_h)$. Distinct $z_h$ guarantee first-factor independence,
but the full repair interpolation is singular. The theorem concerns
these consecutive powers; other one-parameter curves remain possible,
as the explicit instance at the end of this section illustrates.

\begin{theorem}[Consecutive-power interpolation obstruction]
\label{thm:curveobstruction}
Let $t\geq3$, $r\geq2$, $d=tr$, and $m=(t-1)(r-1)+1$.
Over any field, choose distinct $a_1,\ldots,a_d$ and arbitrary
$z_1,\ldots,z_d$. If
\[
x_h=(1,z_h,\ldots,z_h^{t-1}),\qquad
y_h=(1,a_h,\ldots,a_h^{m-1}),
\]
then the $d\beta$-square full interpolation matrix satisfies
\begin{equation}\label{eq:curvecorank}
\operatorname{corank}\mathcal E_H
\geq \binom{(t-2)r}{t-1}>0.
\end{equation}
This holds even when all $z_h$ are distinct.
\end{theorem}
\begin{proof}
Write $s=t-1$ and $L=(t-1)r-1=m+s-2$. Consider
\[
\begin{split}
\mathcal V=\{f(A)+Zg(A):{}&\deg f,\deg g\leq L,\\
&f(a_h)+z_hg(a_h)=0\text{ for all }h\}.
\end{split}
\]
There are $2(L+1)$ coefficients and at most $d$ independent conditions,
so $\ell=\dim\mathcal V\geq(t-2)r\geq s$.
The map $f+Zg\mapsto g$ is injective on $\mathcal V$: if $g=0$, the
polynomial $f$ has $d$ distinct roots and degree at most $L<d$, hence
is zero. Choose a basis $F_i=f_i+Zg_i$, $1\leq i\leq\ell$.
The $g_i$ are linearly independent polynomials.

For each $s$-element subset $I$ define
\begin{equation}\label{eq:alternantkernel}
Q_I(U_1,\ldots,U_s,Z)=
\frac{\det(F_i(U_j,Z))_{i\in I,\,1\leq j\leq s}}
{\prod_{j<k}(U_k-U_j)}.
\end{equation}
Two equal $U$ variables make two determinant columns equal, so every
factor of the denominator divides the numerator. These distinct linear
factors are pairwise coprime in the polynomial ring, and their product
divides it. This reasoning holds also in characteristic two. Swapping
two variables changes the signs of numerator and denominator together;
the quotient is symmetric. Its degree in each $U_j$ is at most
$L-(s-1)=m-1$, and its degree in $Z$ is at most $s$.
It therefore has exactly the form of an admissible order-$s$ unknown:
$\sum_{i=0}^sZ^iQ_{I,i}(U_1,\ldots,U_s)$.

At $U_1=a_h,Z=z_h$, the numerator has a zero first column.
The specialized denominator is a nonzero polynomial in the remaining
independent variables. Since their polynomial ring is a domain,
\begin{equation}\label{eq:invisibleinterpolation}
Q_I(a_h,U_2,\ldots,U_s,z_h)=0\quad\text{for every }h.
\end{equation}
Thus its coefficient array lies in $\ker\mathcal E_H$.

These arrays are nonzero and independent. Their $Z^s$ coefficients are
$\det(g_i(U_j))_{i\in I,j}/\prod_{j<k}(U_k-U_j)$.
To check independence, form the full-column-rank coefficient matrix
$G=[g_1\ \cdots\ g_\ell]$ and choose a left inverse $B$.
Multiply a proposed relation by the common denominator and apply $B$
to each variable's coefficient index. The determinants become the
alternating sums of distinct standard-coordinate tuples indexed by
$I$. Comparing an increasing tuple reads precisely its coefficient,
with sign $1$, so all relation coefficients vanish. Selecting
$s$-element subsets from a basis of $\mathcal V$ produces
$\binom{\ell}{s}$ independent null vectors; since $\ell\geq(t-2)r$ and
$s=t-1$, this yields \eqref{eq:curvecorank}.
\end{proof}

\begin{proposition}[Necessity of $\A_H$ invertibility for single-node repair]\label{prop:repairnecessity}
Assume the node independence hypotheses of Proposition~\ref{prop:code}
(any $t$ of the $x_h$ are independent and any $m$ of the $y_h$ are
independent). Fix a $d$-helper set $H$ and a single failed node
$f\notin H$. Under the linear download rule
\eqref{eq:downloadcoefficients}, node $f$ is exactly repairable from $H$
if and only if $\A_H$ (equivalently $\mathcal E_H$) is invertible.
\end{proposition}
\begin{proof}
If $\A_H$ is invertible, then $\operatorname{im}\A_H=X\otimes\Sym^{t-1}Y$,
so \eqref{eq:targetcriterion} holds and $f$ is exactly repairable from
$H$. Conversely, suppose $f$ is exactly repairable from $H$ under
\eqref{eq:downloadcoefficients}, i.e., $V_f\subseteq\sum_{h\in H}W_h$,
where $W_h:=x_h\otimes y_hy_f\Sym^{t-2}Y\subseteq V_h$ is the
$\beta$-dimensional download subspace from helper $h$. Let $K\subset H$
be any $(k-1)$-subset and set $V_K:=\sum_{i\in K}V_i$. Because
$|K\cup\{f\}|=k$, the file-reconstruction argument in
Proposition~\ref{prop:code} gives $V_f+V_K=X\otimes\Sym^tY$. Comparing
dimensions ($\dim V_h=\alpha$ and $\dim(X\otimes\Sym^tY)=k\alpha$) shows
that this sum is direct; in particular $V_f\cap V_K=0$, so the image
$\overline V_f$ of $V_f$ in the quotient $(X\otimes\Sym^tY)/V_K$ has
dimension $\alpha$. Since $W_i\subseteq V_i\subseteq V_K$ for every
$i\in K$, projecting $V_f\subseteq\sum_{h\in H}W_h$ modulo $V_K$ yields
$\overline V_f\subseteq\sum_{h\in H\setminus K}\overline W_h$, whence
\[
\begin{aligned}
\alpha=\dim\overline V_f
&\leq\dim\sum_{h\in H\setminus K}\overline W_h\\
&\leq\sum_{h\in H\setminus K}\dim W_h
=|H\setminus K|\beta=r\beta=\alpha.
\end{aligned}
\]
Equality forces the subspaces $\overline W_h$ ($h\in H\setminus K$) to
form a direct sum in the quotient and each projection
$W_h\to\overline W_h$ to be injective. Consequently, any linear relation
$\sum_{h\in H}w_h=0$ ($w_h\in W_h$) reduces modulo $V_K$ to
$\sum_{h\in H\setminus K}\overline w_h=0$ and thus forces $w_h=0$ for
all $h\in H\setminus K$. Varying $K$ over all $(k-1)$-subsets of $H$
(noting $k-1<d=|H|$) forces $w_h=0$ for every $h\in H$. Hence
$\dim\sum_{h\in H}W_h=d\beta$. Since
$\sum_{h\in H}W_h=y_f\operatorname{im}\A_H$ and multiplication by the
nonzero form $y_f$ is injective, $\dim\operatorname{im}\A_H=d\beta=t\alpha$,
so the square matrix $\A_H$ (and $\mathcal E_H=\A_H^{\mathsf T}$) is
invertible.
\end{proof}

\begin{corollary}[Repair impossibility on consecutive-power curves]\label{cor:curveimpossibility}
Let $t\geq3$, $r\geq2$, $n\geq d+1$, and $d=tr$. If $a_1,\ldots,a_n$ are
pairwise distinct and $z_1,\ldots,z_n$ are pairwise distinct, then for
$x_h=(1,z_h,\ldots,z_h^{t-1})$ and $y_h=(1,a_h,\ldots,a_h^{m-1})$, no
$d$-helper set $H$ can exactly repair any failed node $f\notin H$ under
the linear download rule \eqref{eq:downloadcoefficients}.
\end{corollary}
\begin{proof}
Distinct $z_h$ and distinct $a_h$ give Vandermonde independence of every
$t$ first factors and every $m$ second factors, so
Proposition~\ref{prop:repairnecessity} applies. Theorem~\ref{thm:curveobstruction}
shows that $\mathcal E_H$ is singular for every $d$-helper set $H$,
ruling out exact repair of every $f\notin H$ under
\eqref{eq:downloadcoefficients}.
\end{proof}

Proposition~\ref{prop:repairnecessity} and
Corollary~\ref{cor:curveimpossibility} apply to the linear download rule
\eqref{eq:downloadcoefficients}. The following example illustrates the
obstruction concretely by exhibiting a kernel element of $\mathcal E_H$
together with a failed-node coefficient that cannot be recovered from
the downloads.
\begin{example}[A target-visible null polynomial]\label{ex:curvefailure}
Over $F=\F_{101}$ take $t=3,r=2$, so
$(n,k,d,\alpha,\beta)=(7,5,6,6,3)$. Let $a_h=h$ and $z_h=h^3$ for
$1\leq h\leq7$. The $z_h$ are $1,8,27,64,24,14,40$, all distinct.
Consequently every three $x_h$ and every three $y_h$ are independent.
Use helpers $1,\ldots,6$, with failed node $7$.

All following coefficients are in $\F_{101}$. The helper polynomial is
\[
\begin{split}
p_H(A)={}&A^6+80A^5+74A^4+73A^3\\
&\quad+8A^2+54A+13.
\end{split}
\]
Set
\[
\begin{aligned}
F_1(A,Z)&=Z-A^3,\\
F_2(A,Z)&=73A^3+8A^2+54A+13\\
&\quad+Z(A^3+80A^2+74A).
\end{aligned}
\]
Here $F_2(A,A^3)=p_H(A)$, so both forms vanish at every helper pair.
The polynomial
\begin{equation}\label{eq:explicitbezoutkernel}
Q(U,V,Z)=\frac{F_1(U,Z)F_2(V,Z)-F_2(U,Z)F_1(V,Z)}{U-V}
\end{equation}
is symmetric in $U,V$, has degree at most two in each variable, and
satisfies $Q(a_h,V,z_h)=0$ for all helpers. Nevertheless,
\[
[U^2V^2]Q(U,V,40)=-(80\cdot40+8)=24\ne0.
\]
The downloaded coefficient data vanish, whereas the failed-node target
does not. These order-two coefficients are realizable by an order-three
file contracted at $y_7$: in a basis beginning with $y_7$, assign the
file entries with one first-basis index to the desired order-two array.
Each array entry has a distinct such file entry, so the contraction is
surjective and the displayed ambiguity arises from an actual file; an
exact matrix check gives rank $17$ for the $18$-square repair matrix.
\end{example}

\noindent\emph{Nonconsecutive one-parameter curves.}
Other monomial exponents avoid the obstruction in
Theorem~\ref{thm:curveobstruction}. Over $\F_{101}$, take $t=3,r=2$ and
\[
\begin{aligned}
a_h&=h,\qquad y_h=(1,a_h,a_h^2),\\
x_h&=(1,a_h^2,a_h^5),\qquad 1\leq h\leq7.
\end{aligned}
\]
Every three $x_h$ and every three $y_h$ are independent. Ordering
helpers increasingly, rows by $(h,\gamma)$ with $\gamma=e_0,e_1,e_2$,
and columns by $i=0,1,2$ and
$(z_0^2,z_0z_1,z_0z_2,z_1^2,z_1z_2,z_2^2)$, the seven determinants of
\eqref{eq:interpolationentry} indexed by the failed node $f=1,\ldots,7$ are
\[
(46,23,38,60,70,8,84)\pmod{101},
\]
all nonzero, certifying an exact-repair code via Proposition~\ref{prop:code}.

\section{Matched Comparisons and Computational Scope}\label{sec:comparison}
Complementing the general scaling comparison in
Table~\ref{tab:guarantees}, Table~\ref{tab:intro} evaluates concrete
MSR code instances at shortest length $n=tr+1$ ($d=n-1$), matching
$(n,k,d)$, storage rate $k/n$, and normalized repair bandwidth $d/(kr)$.
Throughout this section $r=d-k+1$, which can differ from $n-k$ once
$n>d+1$.

\begin{roadmaptable}
\centering\small
\caption{Matched parameters and subpacketizations at shortest length
$n=tr+1$ ($r=d-k+1$). All compared columns share $(n,k,d)$ and
$\beta=\alpha/r$.}
\label{tab:intro}
\begin{tabular}{ccrrrr}
\toprule
$(t,r)$&$(n,k,d)$&$(\alpha,\beta)$ here&$\alpha_{\rm LWHY}$&$\alpha_{\rm YB}$&$\alpha_{\rm LTS}$\\
\midrule
$(3,2)$&$(7,5,6)$&$(6,3)$&8&16&16\\
$(3,3)$&$(10,7,9)$&$(15,5)$&27&81&243\\
$(3,4)$&$(13,9,12)$&$(28,7)$&64&256&16384\\
$(3,72)$&$(217,145,216)$&$(10296,143)$&373248&26873856&$72^{109}$\\
$(4,2)$&$(9,7,8)$&$(20,10)$&8&32&32\\
$(4,3)$&$(13,10,12)$&$(84,28)$&81&243&2187\\
$(4,4)$&$(17,13,16)$&$(220,55)$&256&1024&262144\\
$(4,14)$&$(57,43,56)$&$(11480,820)$&38416&537824&$14^{29}$\\
$(5,2)$&$(11,9,10)$&$(70,35)$&16&64&64\\
$(6,2)$&$(13,11,12)$&$(252,126)$&32&128&128\\
\bottomrule
\end{tabular}
\end{roadmaptable}

The all-node non-optimal-access construction of Li
et~al.~\cite{lwhy} has
\begin{equation}\label{eq:nabenchmark}
\alpha_{\rm NA}=r^{\lceil n/(r+1)\rceil}.
\end{equation}
At $n=tr+1$, this equals $r^3$ for $t=3$, and $r(2r-1)<r^3$ holds for
every $r\geq2$. For $t=4$, $\alpha_{\rm NA}=r^4$ for $r\geq3$, so
$\binom{3r}{3}<\alpha_{\rm NA}$ holds for all $r\geq4$, whereas
$\alpha_{\rm NA}$ is smaller at $(4,2),(4,3),(5,2),(6,2)$ at shortest
length ($8<20$, $81<84$, $16<70$, and $32<252$, where~\cite{wlhy} also
gives $2^{\lceil n/3\rceil}$ at $r=2$). In Table~\ref{tab:intro}, the
$(3,72)$ and $(4,14)$ rows are the first instances exceeding the
$\alpha\leq10011$ computational verification range of~\cite{atrahasis};
asymptotically as $r\to\infty$, our subpacketizations grow as $2r^2$ and
$(9/2)r^3$ versus $r^3$ and $r^4$ for~\cite{lwhy}.

\emph{Length-independent subpacketization for $n>d+1$.}
When $(k,d)$ (and hence $t,r$) are held fixed and the code length $n$
grows, $\alpha$ in \eqref{eq:alphabeta} remains constant and the field
thresholds in \eqref{eq:mainbounds} and \eqref{eq:higherD} grow
polynomially as $O_{t,r}(n^{tr-1})$. By contrast,
$\alpha_{\rm NA}=r^{\lceil n/(r+1)\rceil}$ grows exponentially in $n$.
Consequently, even in the small-$r$ cases $(t,r)=(4,2),(4,3),(5,2),(6,2)$
where $\alpha_{\rm NA}$ is smaller at $n=d+1$, our constant
subpacketizations overtake~\cite{wlhy,lwhy} for every $n\geq13$
($20<32$), $n\geq17$ ($84<243$), $n\geq19$ ($70<128$), and $n\geq22$
($252<256$), respectively. Moreover, because the union-bound field
thresholds grow polynomially as $O_{t,r}(n^{tr-1})$, the per-node bit
granularity $\alpha\log_2 q$ grows only logarithmically in $n$ as
$O_{t,r}(\alpha\log_2 n)$, whereas exponential-in-$n$ constructions incur
exponentially growing $\alpha\log_2 q$. Two structural limits remain.
For fixed $c=n-d$, the storage rate along $d=tr$ satisfies
\[
\frac{k}{n}-\frac{t-1}{t}=\frac{t-(t-1)c}{tn},
\]
which slightly exceeds $(t-1)/t$ at $c=1$ (e.g., $10/13>3/4$ at
$(13,10,12)$) and approaches $(t-1)/t$ as $n$ grows. In addition, dense
Vandermonde vectors $y_f$ require full-symbol helper reads
(Proposition~\ref{prop:access}).

For explicit field bounds at shortest length, put
\begin{equation}\label{eq:lwhyfield}
\widehat n=(r+1)\Bigl\lceil\frac n{r+1}\Bigr\rceil,\qquad
Q_{\rm LWHY}=\widehat n r+r2^{r-1}.
\end{equation}
The non-optimal-access family of~\cite{lwhy} exists over every
prime-power field satisfying the sufficient condition
$q\geq Q_{\rm LWHY}$ at length $\widehat n$ (an alternative
sufficient bound for large $n$ relative to $r$ is given
in~\cite[Lemma~8]{lwhy}), and shortening $\widehat n-n$ information nodes
preserves $\alpha_{\rm NA}$. When $(k,d)$ are fixed and $n$ grows, our
$\alpha$ is constant and $q$ grows polynomially in $n$, whereas
$\alpha_{\rm NA}$ grows exponentially in $n$. Along the shortest line
$n=tr+1$ with fixed $t$ and growing $r$, $D_3$ and $D_a$ in
\eqref{eq:shortfields} are polynomial in $r$, whereas the $r2^{r-1}$
term in $Q_{\rm LWHY}$ grows exponentially.

Li, Tang, and Song (LTS)~\cite[Constr.~1, Thm.~3]{lts2026} support all
$k+1\leq d<n$ with $\alpha_{\rm LTS}=r^{\lceil n/2\rceil}$ under
\begin{equation}\label{eq:ltsfield}
q>r\lceil n/2\rceil,\qquad \gcd(r,q-1)=1.
\end{equation}
For every $t\in\{3,4\}$, $r\geq2$, and $n\geq tr+1$,
\begin{equation}\label{eq:ltscomparison}
\alpha<\alpha_{\rm LTS},
\end{equation}
since $r(2r-1)<r^3<r^4\leq r^{\lceil(3r+1)/2\rceil}$ for $t=3$,
$20<32$ at $(4,2)$, and $\binom{3r}{3}<\tfrac92r^3<r^5\leq r^{2r+1}$ for
$t=4,r\geq3$, while LTS retains broader $(k,d)$ coverage and closed-form
linear-field selection under \eqref{eq:ltsfield}.

As shown in Table~\ref{tab:guarantees}, the Ye--Barg
construction~\cite{ybaccess} uses $q\geq r\lceil n/r\rceil$ and achieves
optimal access at $d=n-1$, for which the lower bound of~\cite{bvk} requires
$\alpha\geq r^t$; our non-optimal-access codes lie strictly
below $r^t$ for all $r\geq2$ at $t=3$ and all $r\geq4$ at $t=4$. The
general exponential lower bound of~\cite{ag} governs unbounded
$k/(d-k+1)$, whereas $k/r\approx t-1$ is bounded along each fixed-$t$
family. For $t\in\{3,4\}$, shortening WTB~\cite{wtb} deletes $t-2$
systematic nodes from a code of length $(r+1)(t-1)+r$ (giving
$\alpha=r^{t-1}$; in general $\alpha=r^{\lceil k/(r+1)\rceil}$) and
repairs systematic nodes only, while product-matrix MSR codes~\cite{pm}
cover $d\geq2k-2$ with $\alpha=r$, complementing the higher-rate
ranges $d=3(k-1)/2$ and $d=4(k-1)/3$ realized here.

\emph{Shortened practical regimes and rate boundaries.}
Near $d=tr$, our codes yield competitive
subpacketization: for $(n,k,d)=(14,10,12)$ ($t=4,r=3$ at $n=d+2$),
Theorem~\ref{thm:main} gives $\alpha=84$ versus $\alpha_{\rm NA}=81$
in~\cite{lwhy} (and $84<243$ once $n\geq17$). Under heavier shortening,
however, Corollary~\ref{cor:shorteneddegree} retains the unshortened
subpacketization $\binom{(t-1)r}{t-1}$: for $(14,10,13)$ ($r=4$,
shortened by $s=3$ from $(17,13,16)$), Corollary~\ref{cor:shorteneddegree}
uses $\alpha=220$ whereas~\cite{lwhy} achieves
$\alpha_{\rm NA}=4^{\lceil14/5\rceil}=64$ (by comparison, cascade
codes~\cite{elyasimohajer2020} have length-independent
$\alpha=r^k=4^{10}$ over $O(n)$ fields). For $r\geq3$, parameters
satisfying $3d<4(k-1)$, such as $(n,k,d)=(20,16,19)$ (where
$3d=57<60=4(k-1)$), lie outside the families established here, where
exponential-in-$n/r$ constructions~\cite{ybaccess,lwhy,lts2026} remain
the applicable choice. At $r=2$, Corollary~\ref{cor:highercode} also
covers the $t=5,6$ MSR code families and their shortened versions.

\subsection{Access, decoding arithmetic, and stripe size}
Access counts depend on the stored basis; the following proposition
evaluates helper reads and operations in the monomial basis of
Proposition~\ref{prop:code}.
\begin{proposition}[Helper access in the monomial basis]\label{prop:access}
If the failed vector $y_f$ has $s_f$ nonzero coordinates in a fixed
basis of $Y$, direct computation of the $\beta$ downloads reads
\begin{equation}\label{eq:access}
A_{\rm help}(s_f)=\alpha-\binom{m-s_f+t-2}{t-1}
\end{equation}
distinct stored symbols, using at most $s_f\beta$ multiplications and
$(s_f-1)\beta$ additions, where $\binom ab=0$ for $0\leq a<b$.
\end{proposition}
\begin{proof}
For each degree-$(t-2)$ monomial $q$, substituting
$y_f=\sum_j\gamma_jz_j$ forms a linear combination of stored coordinates
indexed by $z_jq$ with $\gamma_j\ne0$. Across all $q$, the accessed
coordinates are the degree-$(t-1)$ monomials involving at least one active
variable, whose count is \eqref{eq:access}. Each of the $\beta$ outputs
combines at most $s_f$ stored symbols.
\end{proof}
For dense $y_f$ ($s_f=m$), each helper reads all $\alpha$ stored symbols
and transmits $\beta=\alpha/r$, reading $1/k$ and sending $1/(kr)$ of the
file. At $(t,r)=(3,3)$, for instance, each helper reads $15$ symbols to
send $5$, using at most $25$ multiplications and $20$ additions.

\emph{Replacement-node repair and file reconstruction.}
At the replacement node, Proposition~\ref{prop:subsetrepair} reduces
repair decoding from order $t\alpha$ to the $\alpha\times\alpha$ system
$R$ in \eqref{eq:subsetrepair}. Factoring $R$ for a given $d$-helper set
$H$ takes $O(\alpha^3)$ field operations once per failure pattern. For
each stripe, forming the right-hand side from the $d\beta$ downloaded
symbols, solving the factored $\alpha\times\alpha$ system, and evaluating
\eqref{eq:subsetinterpolation} at $U=x_f$ take $O(\alpha^2+m\alpha)$
field operations ($O(r^4)$ for $t=3$ and $O(r^6)$ for $t=4$).
Reconstructing the file from $k$ nodes via Lemma~\ref{lem:subsetinterpolation}
solves $\binom kt$ linear systems of constant order $t$ and expands the
symmetric basis in $O(\binom kt+k\alpha^2)$ operations for fixed $t$.

\emph{Stripe size in bits.}
Over $\F_{2^b}$, each node stores $\alpha b$ bits per stripe; field order
and subpacketization jointly determine minimum stripe granularity. Using
the sufficient binary fields from \eqref{eq:shortfields} and
\eqref{eq:lwhyfield}, at $(t,r)=(3,3)$ our code stores
$15\cdot7=105$ bits per node versus $27\cdot6=162$ bits for~\cite{lwhy},
whereas at $(4,4)$ the comparison reverses to $220\cdot13=2860$ bits
versus $256\cdot7=1792$ bits.

\section{Conclusion}\label{sec:conclusion}
We proved that the symmetric-algebra MSR repair determinant is a nonzero
polynomial in every characteristic for $t=3,4$ and every integer
$r\geq2$, as well as for $t=5,6$ at $r=2$. This resolves Open Problem~5
of Duursma and Wang~\cite{atrahasis} for $t=3$ and
$(t,r)=(4,2),(5,2),(6,2)$ with arbitrary prescribed MDS second factors,
and for $t=4$ and all $r\geq2$ via joint Vandermonde selection. Every
proof follows the same perturbation--syzygy pattern: degenerating along
$X$-coordinate channels (preceded, for $t=4$ and $r\geq3$, by
characteristic-free Hermite clustering of Vandermonde nodes) reduces the
baseline operator to homogeneous multiplication maps whose kernels come
from Koszul syzygies and whose cokernels are quotient symmetric powers;
first-order Schur lifting then certifies nonvanishing. The resulting
codes achieve length-independent subpacketization
$\alpha=\binom{(t-1)r}{t-1}$ and polynomial field-size thresholds in
every characteristic.

Two algebraic obstacles arise when extending this argument to general
$t$. First, even at $r=2$ ($d=k+1$), where baseline blocks have pure
two-form Koszul kernels $\Sym^{t-3}Y$, the first-order map $E_t$
requires parametrizing syzygies among $t-2$ quadratic pair products in
$(t-2)$-dimensional quotients $Y/W_j$ and gluing local degree-$(t-3)$
polynomials across overlaps. Pairwise Koszul syzygies of quadrics and
pairwise overlap conditions still close the system at $t=5,6$
(Appendices~\ref{app:t5}--\ref{app:t6}), but for $t\geq7$ the
$t-2\geq5$ pair products and degree-$(t-3)$ local forms ($t-3\geq4$)
produce higher-order syzygy chains and multi-overlap compatibility
constraints. Second, for $t\geq5$ and general $r\geq3$, single-stage
$X$-channel degeneration leaves quotient spaces $Y/W_i$ of dimension
$(t-2)(r-1)$ coupled with syzygies of degree $t-3\geq2$. Extending the
two-stage Vandermonde Hermite degeneration of Section~\ref{sec:t4}
instead yields $t$-node clusters with $(t-1)$-dimensional jet spaces
$Z_g$ and support-triangular blocks of every support size
$1\leq|S|\leq t-1$. Support-triangular elimination and pure-support
cokernel decoupling carry over to arbitrary $t$, but proving generic
nonvanishing for all multi-support diagonal blocks $2\leq|S|\leq t-1$
uniformly in $t$ remains open, as does determining whether arbitrary
prescribed MDS second factors already suffice starting at
$(t,r)=(4,3)$.

Combining the determinant theorems with individual-degree bounds yields
a single small-field configuration valid across all reconstruction and
helper-set subsets. At fixed excess length $n-d$, it can be found
deterministically in polynomial time, while helper-set interpolation
gives closed-form binary coordinates of polynomial extension degree. It
remains open to construct closed-form node coordinates directly over
$O(n^2)$ and $O(n^4)$ fields without candidate search.

\section*{Acknowledgment}
Generative AI tools were used to assist with language polishing and
auxiliary script checks.

\appendices
\section{Derivation of the Hermite Coordinates}\label{app:hermite}
This appendix supplies the coordinate step in \eqref{eq:jets}.
Define
\[
\begin{aligned}
R(z)&=\prod_g(z-a_g)^3,&R_g(z)&=\frac{R(z)}{(z-a_g)^3},\\
B_g&=R_g(a_g),&T_g(z)&=R_g(z)/B_g.
\end{aligned}
\]
At $z=a_g+u$, write
$T_g=1+S_gu+D_gu^2+O(u^3)$. Direct product expansion gives
$S_g=3\sum_{h\neq g}(a_g-a_h)^{-1}$ over an arbitrary field.
Define
\begin{align}\label{eq:hermitebasis}
H_{g,2}&=T_g(z)(z-a_g)^2,\\
H_{g,1}&=T_g(z)\bigl((z-a_g)-S_g(z-a_g)^2\bigr),\nonumber\\
H_{g,0}&=T_g(z)\bigl(1-S_g(z-a_g)+(S_g^2-D_g)(z-a_g)^2\bigr).\nonumber
\end{align}
At $a_g$, their constant, linear, and quadratic coefficients are
$(0,0,1)$, $(0,1,0)$, and $(1,0,0)$, respectively, and each vanishes to
order three at every other center while having degree less than $3G$.
Together with the monic $R$ they form a basis of polynomials of degree
at most $3G$: zero local data imply divisibility by $R$, and a zero
leading coefficient then implies that the polynomial vanishes.

For $p(z)=\sum_{j=0}^{3G}p_jz^j$, its coefficient row defines a linear
functional on $Y$ satisfying $p\cdot y(a)=p(a)$, so a basis of
coefficient rows specifies an invertible coordinate change on $Y$.
Use coordinate rows $\eps^{-2}R$ and $B_g\eps^{-c}H_{g,c}$, $c=0,1,2$,
and scale each central node by $\eps^2/R(A_i)$ and each local node by
$B_g^{-1}$; all operations are invertible over $K((\eps))$.

At a central node, the $f$ coordinate is one and
\[
\begin{aligned}
H_{g,2}&=\frac{R}{B_g(z-a_g)},\\
H_{g,1}&=\frac{R}{B_g}\left((z-a_g)^{-2}-S_g(z-a_g)^{-1}\right).
\end{aligned}
\]
Its $w_g$ coordinate is therefore $c_{g,i}$, its $z_{g,1}$ coordinate is
$\eps(c_{g,i}^2-S_gc_{g,i})$, and its $z_{g,0}$ coordinate has order
at least two. This proves the central formula in \eqref{eq:jets}.

At a local node $a_g+\eps b_i$, we have
$R=B_g\eps^3b_i^3+O(\eps^4)$.
Its $f$ coordinate is $\eps b_i^3+O(\eps^2)$, and
\eqref{eq:hermitebasis} gives its own group coordinates
\[
1+O(\eps^3),\qquad b_i+O(\eps^2),\qquad
b_i^2+\eps S_gb_i^3+O(\eps^2).
\]
For $h\neq g$, the identity $H_{h,2}=R/[B_h(z-a_h)]$ gives the
$w_h$ coordinate $\eps b_i^3/(a_g-a_h)+O(\eps^2)$.
The other coordinates of group $h$ have order at least two.
Summing these terms proves the local formula in \eqref{eq:jets}.
When $G=1$, $T_g=1$ and $S_g=D_g=0$, which also proves the exact
one-group formulas \eqref{eq:singleexact}.

\section{A Complete Eight-Dimensional Lifting Certificate}\label{app:liftcertificate}
This appendix makes the $r=3$ running case reproducible over $\F_{101}$.
It is a certificate for the normalized degeneration, with $\eps$ formal;
the all-characteristic theorem is proved in Section~\ref{sec:t4}.
Use the ordered basis
\[
(f,z_{1,0},z_{1,1},w_1,z_{2,0},z_{2,1},w_2)
\]
and the scalar parameters
\[
(A_0,A_1,A_2,A_3)=(0,1,2,3),\quad(a_1,a_2)=(4,5),\quad b_i=i.
\]
The central first factors are $x_{0,i}=e_i$. The rows of the following
matrices are the local first factors $x_{g,i}^{\mathsf T}$:
\[
X_1=\begin{pmatrix}18&3&29&25\\27&43&38&27\\44&59&7&1\\95&45&37&36\end{pmatrix},\quad X_2=\begin{pmatrix}61&90&15&65\\93&51&60&22\\16&38&84&85\\68&74&69&17\end{pmatrix}.
\]
The zero- and first-order node coefficients are specified by \eqref{eq:jets}.
Form $\A_0,\A_1$ using ordinary quadratic/cubic monomials, ordered as
nondecreasing tuples of basis indices in lexicographic order. Order
source columns first by central nodes $i=0,\ldots,3$, then local group
$1$ and group $2$, with the same node order in each; order target rows
first by $X$ coordinate, then by cubic monomial.

For a source vector $q$, take the eight anchor coefficients, in order,
\begin{equation}\label{eq:certificateanchors}
\psi(q)=\Bigl([fw_g]q_{g,3},[z_{g,0}w_g]q_{g,3},
[z_{g,1}w_g]q_{g,3},[w_g^2]q_{g,3}\Bigr)_{g=1,2}.
\end{equation}
Exact elimination gives singleton ranks $72,72$, double-support rank
$180$, and $\rank\A_0=328$. Restriction of $\psi$ to $\ker\A_0$
is invertible. Thus the $j$th column of $N_0$ is uniquely specified by
\[
\A_0N_0=0,\qquad \psi(N_0)=I_8.
\]
Order the rows of $\Phi$ as $(\Phi_{1,0},\ldots,\Phi_{1,3},
\Phi_{2,0},\ldots,\Phi_{2,3})$, using \eqref{eq:globalcoker} and the
dual bases of the columns of $X_g^{\mathsf T}$. The resulting map over $\F_{101}$ is
\begin{equation}\label{eq:certificate8}
\Phi\A_1N_0=
\left(\begin{array}{rrrr|rrrr}
81&0&12&90&0&0&0&0\\
61&73&46&41&0&0&0&0\\
27&10&53&64&0&0&0&0\\
42&8&65&74&0&0&0&0\\
0&0&0&0&56&0&61&20\\
0&0&0&0&69&44&18&16\\
0&0&0&0&59&5&57&14\\
0&0&0&0&66&19&72&73
\end{array}\right).
\end{equation}
Its two diagonal blocks have determinants $74$ and $6$, so the full
determinant is $74\cdot6=40\ne0$ in $\F_{101}$.
Recomputing after deleting the local first-order motions and the
central $-S_gc_{g,i}$ terms gives the identical matrix, and applying
the surviving motion to the higher-support part of $N_0$ gives zero
under every row of $\Phi$, illustrating both assertions of
Lemma~\ref{lem:puresupportlift}.

\section{Deterministic Selection from Local Nonzero Evaluations}\label{app:selector}

\begin{lemma}[Common specialization]\label{lem:commonwitness}
For each polynomial $f$ in a finite list over $\F_q$, suppose a
point $w_f$ with $f(w_f)\ne0$ and an evaluation algorithm are given.
Put $D_j=\sum_f\deg_{z_j}f$. If $q>\max_j D_j$, a simultaneous nonzero
point can be constructed deterministically. For coordinate $j$, at most
$D_j+1$ candidate values need be tested against the factors involving $j$.
\end{lemma}
\begin{proof}
Process coordinates in a fixed order. At the start of a step all
$w_f$ agree on the prefix already chosen. Replacing coordinate $j$ in
$w_f$ by $T$ gives a univariate polynomial that is nonzero at the
old coordinate. Its roots therefore exclude at most $\deg_{z_j}f$
values. The union excludes at most $D_j$. Choose the first admissible
value among $D_j+1$ distinct field elements and update coordinate $j$
in every witness. This preserves the invariant and extends the common
prefix. The last update gives the desired point. Upper bounds on the
degrees suffice throughout.
\end{proof}

\begin{proof}[Proof of Theorem~\ref{thm:detselection}]
Apply Lemma~\ref{lem:commonwitness} to the factors of $P_t$.
Each first-factor minor has an affine Vandermonde basis as a witness,
and each pair difference has the witness $(0,1)$. It remains to
initialize the repair determinants in $\F_q$.

\emph{Fixed-MDS witnesses for $t=3$ and $(t,r)=(4,2)$.}
For $t=3$, each helper set uses its prescribed $y_h$ and the pencil
\eqref{eq:t3pencil} with $r$ distinct $\lambda_a$; by
\eqref{eq:t3leadingcertificate}, $\det\A_H(s)$ is nonzero of degree at
most $d\beta$ in $s$, so testing $d\beta+1\leq D_3+1$ values yields a
nonzero evaluation. Similarly, at $(t,r)=(4,2)$ with prescribed MDS
$y_h$ and $q>D_x$, the pencil \eqref{eq:singleeta} in
Proposition~\ref{prop:t4r2} has degree at most $4\beta=40<D_x$ in $\eta$,
so testing $41$ values initializes each helper set.

\emph{A $t=4$ witness shared by relabeling.}
For joint Vandermonde selection, fix one ordered template of four
$A_i$ and $G=r-1$ centers $a_g$ mutually distinct, together with four
distinct $b_i$. Impose on unrestricted first-factor coordinates:
\begin{itemize}
\item the $r$ basis determinants of the central and local four-tuples;
\item for each singleton, $[\eps^4]\det M_g(\eps)$ in \eqref{eq:singleexact}
(evaluated division-free modulo $\eps^5$);
\item the normalized zero-order double- and triple-support diagonal
determinants of Table~\ref{tab:blocks}.
\end{itemize}
Using the standard basis, \eqref{eq:singleeta} (degree $\leq80$), and
\eqref{eq:doubleeta} with distinct $\lambda,\mu\notin\{0,1\}$ (degrees
$\leq180$ and $108$), at most $181\leq q$ evaluations initialize all
local constraints. Their total-degree sum is
\[
4r+80G+180\binom G2+108\binom G3=4\alpha+8G.
\]
Since $4\alpha+8G<12(r-1)\alpha=(m-1)d\beta<q$, a preliminary pass of
Lemma~\ref{lem:commonwitness} makes all local
witnesses simultaneous. Proposition~\ref{prop:globalassembly} then makes
the original Vandermonde determinant on $A_i,a_g+\eps b_i$ a nonzero
polynomial in $\eps$ of degree at most $(m-1)d\beta<q$; testing
$(m-1)d\beta+1$ values yields a numerical repair witness shared by
relabeling across all helper sets.

\emph{Normalization and operation count.}
For each repair witness, choose $b\in\F_q$ such that $(1,b,\ldots,b^{t-1})$
is nonzero on all $d$ first factors (at most $d(t-1)<q$ values are
excluded), complete this row to an invertible basis change on $X$, and
rescale every first factor to leading coordinate one. Now
Lemma~\ref{lem:commonwitness} applies with individual-degree bounds
\eqref{eq:D3} and \eqref{eq:D4}.

With $M=\binom nd$ and $N=t\alpha$, direct candidate testing evaluates
at most $D_j+1$ values on the $\binom{n-1}{d-1}$ repair matrices
involving node $i$, taking $O(n D_{\max}\binom{n-1}{d-1}N^3+n\binom{n-1}{t-1}D_{\max})$
field operations. Alternatively, since each coordinate $T$ enters one
repair matrix with degree $e\leq\beta$ (for $x$ coordinates) or
$e\leq(m-1)\beta$ (for $a$ coordinates), interpolating $\det\A_H(T)$ from
$e+1$ determinant evaluations and evaluating the resulting polynomial on
$D_j+1$ points replaces $D_{\max}$ in the matrix eliminations by $e+1$,
giving elimination cost $O(n\binom{n-1}{d-1}(e+1)N^3)$ plus
$O(n\binom{n-1}{d-1}(e+1)(e+D_{\max})+n\binom{n-1}{t-1}D_{\max})$ scalar
operations. At $n=d+1$ (where $\binom{n-1}{d-1}=n-1$), this yields
$O(n^9)$ field operations for $t=3$ ($e=O(n)$, $N=\Theta(n^2)$) and
$O(n^{14})$ for $t=4$ ($e=O(n^3)$ on $a$ coordinates, $N=\Theta(n^3)$),
and polynomial complexity in $n$ for every fixed $c=n-d$.
\end{proof}

\section{Polynomial Relations and Interpolation for $t=5$, $r=2$}\label{app:t5}
This appendix proves the $t=5$ case of Theorem~\ref{thm:higherrepair}.
The ten second-factor vectors are fixed with every five independent.
Removing $s$ disjoint node pairs and quotienting by their $2s$-dimensional
span leaves every $5-2s$ remaining node images independent, because any
dependence would lift to at most five original MDS vectors; in particular,
in every one-dimensional quotient below, all remaining node images are
nonzero.

\subsection{Two polynomial lemmas}
\begin{lemma}[Three quadratic products]\label{lem:5quadrics}
In a three-variable polynomial ring over a field, let
$Q_1=ab$, $Q_2=cd$, $Q_3=ef$, where every three of the six linear forms
are independent. If quadratic forms $g_1,g_2,g_3$ satisfy
$Q_1g_1+Q_2g_2+Q_3g_3=0$, then unique scalars $A,B,C$ satisfy
\begin{equation}\label{eq:5syz}
\begin{aligned}
g_1&=Q_2A+Q_3B,\\
g_2&=-Q_1A+Q_3C,\\
g_3&=-Q_1B-Q_2C.
\end{aligned}
\end{equation}
\end{lemma}
\begin{proof}
We first verify
\begin{equation}\label{eq:5ideal}
(ab,cd)=(a,c)\cap(a,d)\cap(b,c)\cap(b,d).
\end{equation}
Let $P$ belong to the right side. Modulo $a$, the coprime forms $c,d$
both divide $P$, so $P=cdU+aV$. Modulo $b$, the same reasoning shows
$cd\mid P$, hence $cd\mid aV$. In the quotient by $b$, $a$ is coprime
to each of $c,d$ by three-form independence, so $cd\mid V$ modulo $b$.
Lifting gives $V=cdS+bT$, and $P=cd(U+aS)+abT\in(ab,cd)$. The reverse
inclusion is immediate. The same divisibility argument establishes
\eqref{eq:5ideal} in four variables whenever every relevant triple is
independent.

Each quotient on the right of \eqref{eq:5ideal} is a domain, and the
quotient by $(ab,cd)$ embeds in their direct sum. Neither $e$ nor $f$ is
zero in any summand, so $ef$ is a non-zero-divisor modulo $(ab,cd)$.
Reducing $Q_1g_1+Q_2g_2+Q_3g_3=0$ modulo $(Q_1,Q_2)$ yields
$g_3\in(Q_1,Q_2)$, and degree two gives $g_3=-Q_1B-Q_2C$ for scalars
$B,C$. Substituting gives $Q_1(g_1-Q_3B)+Q_2(g_2-Q_3C)=0$; coprimality
of $Q_1,Q_2$ and degree two give $g_1-Q_3B=Q_2A$ and
$g_2-Q_3C=-Q_1A$ for one scalar $A$. For uniqueness, $g_1=g_2=g_3=0$
forces $B=C=0$ by independence of $Q_1,Q_2$, and then $A=0$.
\end{proof}

\begin{lemma}[Three quotients separate quadratics]\label{lem:5separation}
Let $W_1,W_2,W_3$ be spanned by three disjoint node pairs in
five-dimensional $Y$, with every five of their six vectors independent.
If a quadratic vanishes in each $\Sym^2(Y/W_i)$, it is zero in $\Sym^2Y$.
\end{lemma}
\begin{proof}
Choose a basis $a,b,c,d,e$ with $W_1=\Span(a,b)$, $W_2=\Span(c,d)$, and
$W_3=\Span(e,L)$, where $L=Aa+Bb+Cc+Dd+Ee$. Independence after omitting
each of $a,b,c,d$ implies $A,B,C,D\ne0$. A quadratic in
$(a,b)\cap(c,d)$ is a linear combination of $ac,ad,bc,bd$, so it contains
neither $e$ nor pure $(a,b)$ or $(c,d)$ terms. If it also lies in
$(e,L)$, setting $e=0$ makes it divisible by $L_0=Aa+Bb+Cc+Dd$. Writing
it as $L_0(H_{ab}+H_{cd})$ with $H_{ab}\in\Span(a,b)$ and
$H_{cd}\in\Span(c,d)$, its pure $(a,b)$ part $(Aa+Bb)H_{ab}=0$ forces
$H_{ab}=0$, and similarly $H_{cd}=0$.
\end{proof}

\subsection{Paired baseline and cyclic first-order map}
\begin{proof}[Proof of Theorem~\ref{thm:higherrepair} for $t=5$]
\emph{Step 1: baseline defect.}
Pair the nodes arbitrarily as $F_i,L_i$ ($i\in\mathbb Z/5\mathbb Z$),
with $W_i=\Span(F_i,L_i)$ and $Q_i=F_iL_i$. At $x_{F_i}=x_{L_i}=e_i$,
the order-$350$ repair map splits into five $70\times70$ blocks
\[
\begin{aligned}
\mu_i&:(\Sym^3Y)^2\longrightarrow\Sym^4Y,\\
&(p_i,q_i)\longmapsto F_ip_i+L_iq_i.
\end{aligned}
\]
By Lemma~\ref{lem:koszul}, the kernel and cokernel of $\mu_i$ are
\begin{equation}\label{eq:5base}
\begin{aligned}
\ker\mu_i&=\{(L_i\ell_i,-F_i\ell_i):\ell_i\in\Sym^2Y\},\\
\coker\mu_i&=\Sym^4(Y/W_i),
\end{aligned}
\end{equation}
each of dimension $\binom62=\binom64=15$, giving total defect $75$.

\emph{Step 2: cyclic perturbation.}
Over $F(\lambda)$, set
\begin{equation}\label{eq:5perturb}
\begin{aligned}
x_{F_i}&=e_i+\eta(e_{i-2}+\lambda e_{i-3}+\lambda^2e_{i-4}),\\
x_{L_i}&=e_i.
\end{aligned}
\end{equation}
On \eqref{eq:5base}, the induced $\eta$-first-order map $E_5$ has output
\begin{equation}\label{eq:5E}
\begin{aligned}
(E_5\ell)_j&=\overline Q_{j+2}\overline\ell_{j+2}
+\lambda\overline Q_{j+3}\overline\ell_{j+3}\\
&\quad+\lambda^2\overline Q_{j+4}\overline\ell_{j+4}\in\Sym^4(Y/W_j),
\end{aligned}
\end{equation}
where bars denote reduction modulo $W_j$. By Lemma~\ref{lem:lift}, it
suffices to prove this $75\times75$ map injective.

\emph{Step 3: local syzygy parametrization.}
Suppose $E_5\ell=0$. In $Y/W_j$, every three of the six factors in
\eqref{eq:5E} are independent. Applying Lemma~\ref{lem:5quadrics} yields
unique scalars $A_j,B_j,C_j$ such that
\begin{equation}\label{eq:5local}
\begin{aligned}
\overline\ell_{j+2}&=\overline Q_{j+3}A_j+\overline Q_{j+4}B_j,\\
\lambda\overline\ell_{j+3}&=-\overline Q_{j+2}A_j+\overline Q_{j+4}C_j,\\
\lambda^2\overline\ell_{j+4}&=-\overline Q_{j+2}B_j-\overline Q_{j+3}C_j.
\end{aligned}
\end{equation}

\emph{Step 4: overlap elimination.}
For $j\ne k$, pick a basis $u_{jk}$ of the one-dimensional space
$Y/(W_j+W_k)$ and write $Q_h=q_h^{jk}u_{jk}^2$ modulo $W_j+W_k$. For
$h\notin\{j,k\}$, $q_h^{jk}\ne0$ by five-node independence, so the ratios
\begin{equation}\label{eq:5ratios}
R_j=\frac{q_{j+2}^{j,j+1}}{q_{j+3}^{j,j+1}},\qquad
T_j=\frac{q_{j+1}^{j,j+2}}{q_{j+3}^{j,j+2}}
\end{equation}
are nonzero and basis-independent. Comparing $\ell_{j+4}$ at targets $j$
and $j+2$ modulo $W_j+W_{j+2}$ gives
$-\lambda^{-2}q_{j+3}^{j,j+2}C_j=q_{j+1}^{j,j+2}B_{j+2}$, so
\begin{equation}\label{eq:5C}
C_j=-\lambda^2T_jB_{j+2}.
\end{equation}
Comparing $\ell_{j+4}$ at targets $j$ and $j+1$ modulo $W_j+W_{j+1}$
gives $-q_{j+2}^{j,j+1}B_j-q_{j+3}^{j,j+1}C_j=-\lambda q_{j+3}^{j,j+1}A_{j+1}$,
so
\begin{equation}\label{eq:5A}
A_{j+1}=\lambda^{-1}R_jB_j-\lambda T_jB_{j+2}.
\end{equation}
Comparing $\ell_{j+3}$ at the same two targets gives
\begin{equation}\label{eq:5compat}
q_{j+2}^{j,j+1}A_j-q_{j+4}^{j,j+1}C_j+\lambda q_{j+4}^{j,j+1}A_{j+1}=0.
\end{equation}
Substituting \eqref{eq:5C} and \eqref{eq:5A} cancels the
$\lambda^2B_{j+2}$ terms and leaves
\begin{equation}\label{eq:5cycle}
\lambda^{-1}L_jB_{j-1}+D_jB_j+\lambda U_jB_{j+1}=0,
\end{equation}
where
\begin{equation}\label{eq:5coeff}
\begin{aligned}
L_j&=q_{j+2}^{j,j+1}R_{j-1},\\
D_j&=q_{j+4}^{j,j+1}R_j,\\
U_j&=-q_{j+2}^{j,j+1}T_{j-1}
\end{aligned}
\end{equation}
are nonzero and independent of $\lambda$.

\emph{Step 5: leading Laurent term.}
In the $5\times5$ matrix $M_5(\lambda)$ of \eqref{eq:5cycle}, every
determinant term has Laurent degree at most $5$, achieved uniquely by
choosing $\lambda U_j$ in every row (a positive $5$-cycle):
\begin{equation}\label{eq:5leading}
[\lambda^5]\det M_5(\lambda)=\prod_{j=0}^4U_j\ne0.
\end{equation}
Hence all $B_j=0$ over $F(\lambda)$, forcing $A_j=C_j=0$. By
\eqref{eq:5local} each $\ell_i$ vanishes in its three target quotients,
so Lemma~\ref{lem:5separation} forces $\ell_i=0$. Thus $E_5$ is
invertible, and Lemma~\ref{lem:lift} yields a nonzero $\eta^{75}$ term
in the specialized repair determinant over $F(\lambda,\eta)$, proving
polynomial nonvanishing over $F$ in every characteristic.
\end{proof}

\section{Polynomial Relations and Interpolation for $t=6$, $r=2$}\label{app:t6}
This appendix proves the $t=6$ case of Theorem~\ref{thm:higherrepair}.
The twelve second-factor vectors are fixed with every six independent;
quotienting by $s$ disjoint node-pair planes preserves $(6-2s)$-tuple
independence. Table~\ref{tab:t6dependency} summarizes the reduction
chain $1512\to336\to144\to12$, and Table~\ref{tab:t6quotients} records
the three quotient spaces appearing in the proof.

\begin{roadmaptable}
\centering\small\setlength{\tabcolsep}{5pt}
\caption{Objects and proof dependencies in the $t=6,r=2$ argument.}
\label{tab:t6dependency}
\begin{tabular}{p{0.08\textwidth}p{0.29\textwidth}p{0.25\textwidth}p{0.26\textwidth}}
\toprule
Size&Object counted&Result used&Role in proof\\
\midrule
$1512$&Order of repair matrix ($6\alpha$, $\alpha=252$)&Proposition~\ref{prop:6paired}&Paired baseline splits the map into six blocks.\\
$336$&Baseline kernel/cokernel dimension; order of $E_6$&Proposition~\ref{prop:6paired}; Lemma~\ref{lem:lift}&Invertibility of $E_6$ lifts the baseline.\\
$144$&Local parameters and compatibility rows; order of $C_6$&Lemmas~\ref{lem:6quadrics}--\ref{lem:6gluing}; Proposition~\ref{prop:6equivalence}&Exact kernel isomorphism $\ker E_6\simeq\ker C_6$.\\
$12$&Kernel/cokernel dimension of leading matrix $C_0$ (rank $132$)&Lemmas~\ref{lem:6leading}--\ref{lem:6firstorder}&Six $2$-dimensional isomorphisms lift $C_0$.\\
\bottomrule
\end{tabular}
\end{roadmaptable}

\begin{roadmaptable}
\centering\small\setlength{\tabcolsep}{5pt}
\caption{Three quotient spaces in Appendix~\ref{app:t6}, where $W_i=\Span(F_i,L_i)$ and $U_j=Y/(W_j+W_{j+2})$.}
\label{tab:t6quotients}
\begin{tabular}{p{0.34\textwidth}cp{0.48\textwidth}}
\toprule
Space&Dim.&Role\\
\midrule
$\Sym^5(Y/W_i)$&$56$&Output cokernel of paired baseline block $\mu_i$ in \eqref{eq:6base}.\\
$\Sym^3(Y/W_j)$&$20$&Local restriction of a global cubic $\ell_i$, glued in Lemma~\ref{lem:6gluing}.\\
$H_j=\Sym^3U_j/(Q_B\Sym^1U_j)$&$2$&Cokernel of one leading compatibility block after $g_D$-elimination in \eqref{eq:6smallcokernel}.\\
\bottomrule
\end{tabular}
\end{roadmaptable}

\subsection{Local relations among four quadratic products}
\begin{lemma}[Four quadratic products]\label{lem:6quadrics}
In a four-variable polynomial ring over a field, let $Q_i$, $1\leq i\leq4$,
be products of four disjoint pairs among eight linear forms, with every
four independent. If cubics $g_i$ satisfy $\sum_{i=1}^4Q_ig_i=0$, then
unique linear forms $c_{ik}$ satisfy
\begin{equation}\label{eq:6syz}
g_i=\sum_{k\ne i}Q_kc_{ik},\qquad c_{ki}=-c_{ik},\quad c_{ii}=0.
\end{equation}
\end{lemma}
\begin{proof}
The divisibility proof of \eqref{eq:5ideal} applies in four variables.
For three pair products $ab,cd,ef$, we also have
\begin{equation}\label{eq:6ideal}
(ab,cd,ef)=\bigcap_{\substack{h\in\{a,b\},\ k\in\{c,d\}\\l\in\{e,f\}}}(h,k,l).
\end{equation}
For the nontrivial inclusion, let $P$ belong to the right side.
Modulo $e$, \eqref{eq:5ideal} gives $P=abU+cdV+eT$. Modulo $f$, the same
identity gives $eT\in(ab,cd)$. The quotient by $(f,ab,cd)$ embeds in the
four domains modulo $(f,h,k)$, in each of which $e\ne0$ by four-form
independence; hence $T=abS+cdR+fK$ and $P\in(ab,cd,ef)$.

By \eqref{eq:5ideal} and \eqref{eq:6ideal}, $Q_3$ is a non-zero-divisor
modulo $(Q_1,Q_2)$ and $Q_4$ is a non-zero-divisor modulo
$(Q_1,Q_2,Q_3)$. Reducing $\sum_iQ_ig_i=0$ modulo $(Q_1,Q_2,Q_3)$ gives
$g_4=-\sum_{i=1}^3Q_ic_{i4}$ with linear $c_{i4}$. Substituting and
reducing modulo $(Q_1,Q_2)$, then using coprimality of $Q_1,Q_2$, yields
\eqref{eq:6syz}. For uniqueness, setting all $g_i=0$ forces the linear
coefficient of $Q_3$ in the fourth row into $(Q_1,Q_2)$ and hence to
zero; coprimality kills the remaining linear coefficients in rows four,
three, and two in turn.
\end{proof}

\subsection{Cubic gluing from pairwise overlap consistency}
\begin{lemma}[Cubic gluing]\label{lem:6gluing}
Let $W_1,\ldots,W_4$ be spanned by four disjoint node pairs in
six-dimensional $Y$, with every six of their eight vectors independent.
Cubics $p_j\in\Sym^3(Y/W_j)$ come from a unique
$\ell\in\Sym^3Y$ if and only if their images agree in
$\Sym^3(Y/(W_j+W_k))$ for every $j<k$.
\end{lemma}
\begin{proof}
\emph{Injectivity of joint restriction.}
The first three pairs give $Y=W_1\oplus W_2\oplus W_3$. Writing the
fourth pair $u,v$ in block components, independence forces their two
components in each block to be independent, so after block basis changes
$u=a_1+a_2+a_3$ and $v=b_1+b_2+b_3$. A cubic zero modulo $W_1,W_2,W_3$
is a linear combination of the eight monomials
\begin{equation}\label{eq:6eight}
\begin{aligned}
&a_1a_2a_3,\quad a_1a_2b_3,\quad a_1b_2a_3,\quad a_1b_2b_3,\\
&b_1a_2a_3,\quad b_1a_2b_3,\quad b_1b_2a_3,\quad b_1b_2b_3.
\end{aligned}
\end{equation}
Modulo $W_4$, substituting $a_3=-a_1-a_2$ and $b_3=-b_1-b_2$ isolates
items $1,3,4,5,6,8$ of \eqref{eq:6eight} via the unique monomials
$a_1^2a_2,a_1^2b_2,a_1b_2^2,b_1a_2^2,b_1^2a_2,b_1^2b_2$. Items $2$ and $7$
then give $-a_1a_2b_1-a_1a_2b_2$ and $-a_1b_1b_2-a_2b_1b_2$, whose
monomials are disjoint, forcing all eight coefficients to zero. Thus
$R:\Sym^3Y\to\bigoplus_j\Sym^3(Y/W_j)$ is injective with image dimension
$\binom83=56$.

\emph{Surjectivity of pairwise differences.}
Define
\begin{equation}\label{eq:6D}
\begin{aligned}
D&:\bigoplus_{j=1}^4\Sym^3(Y/W_j)\longrightarrow
\bigoplus_{j<k}\Sym^3(Y/(W_j+W_k)),\\
&D(p)_{jk}=p_j-p_k,
\end{aligned}
\end{equation}
from dimension $4\binom63=80$ to $6\binom43=24$. Fix $j$; in $Y/W_j$ the
other three planes are pairwise complementary, so we may choose bases
with planes $\Span(a_1,a_2)$, $\Span(b_1,b_2)$, and
$\Span(a_1+b_1,a_2+b_2)$. Pure $b$ and pure $a$ cubics independently
prescribe the first two quotients, while modulo the third ($b_h=-a_h$)
the four mixed cubics $a_1b_1^2,a_2b_1^2,a_1b_2^2,a_2b_2^2$ map to
$a_1^3,a_1^2a_2,a_1a_2^2,a_2^3$ without affecting the first two
quotients. Hence summand $j$ can prescribe any single incident edge of
$D$ while vanishing on all other edges, so $D$ is surjective.

\emph{Gluing.}
Since $\operatorname{im}R\subseteq\ker D$ and
$\dim\ker D=80-24=56=\dim\operatorname{im}R$, we have
$\operatorname{im}R=\ker D$, proving existence and uniqueness of gluing.
\end{proof}

\subsection{Paired defect and compatibility system}
\begin{proposition}[Paired $336$-dimensional defect]\label{prop:6paired}
The paired baseline has kernel and cokernel dimension $336$. Under
perturbation \eqref{eq:6perturb}, its induced $\eta$-first-order map is
$E_6$ in \eqref{eq:6E}. If $E_6$ is invertible over $F(\zeta)$ for an
indeterminate $\zeta$, the order-$1512$ repair determinant is nonzero.
\end{proposition}
\begin{proof}
Pair the nodes as $F_i,L_i$ ($i\in\mathbb Z/6\mathbb Z$) with
$W_i=\Span(F_i,L_i)$ and $Q_i=F_iL_i$. At $x_{F_i}=x_{L_i}=e_i$, each
$252\times252$ block $\mu_i:(p_i,q_i)\mapsto F_ip_i+L_iq_i$ has, by
Lemma~\ref{lem:koszul},
\begin{equation}\label{eq:6base}
\begin{aligned}
\ker\mu_i&=\{(L_i\ell_i,-F_i\ell_i):\ell_i\in\Sym^3Y\},\\
\coker\mu_i&=\Sym^5(Y/W_i),
\end{aligned}
\end{equation}
each of dimension $\binom83=\binom85=56$, giving total defect $336$.
Over $F(\zeta)$, choose
\begin{equation}\label{eq:6perturb}
\begin{aligned}
x_{F_i}&=e_i+\eta(\zeta e_{i-2}+e_{i-3}
                 +\zeta^2e_{i-4}+\zeta^3e_{i-5}),\\
x_{L_i}&=e_i.
\end{aligned}
\end{equation}
The induced $\eta$-first-order map is
\begin{equation}\label{eq:6E}
\begin{aligned}
(E_6\ell)_j&=\sum_{i\in S_j}w_{ji}\overline Q_i\overline\ell_i
\in\Sym^5(Y/W_j),\\
S_j&=\{j+2,j+3,j+4,j+5\},
\end{aligned}
\end{equation}
with bars denoting reduction modulo $W_j$ and weights
\begin{equation}\label{eq:6weights}
(w_{j,j+2},w_{j,j+3},w_{j,j+4},w_{j,j+5})=(\zeta,1,\zeta^2,\zeta^3).
\end{equation}
Applying Lemma~\ref{lem:lift} over $F(\zeta)[[\eta]]$ with parameter
$\eta$ and defect $336$ completes the claim.
\end{proof}

\begin{proposition}[Compatibility kernel isomorphism]\label{prop:6equivalence}
Over $F(\zeta)$, $\ker E_6\simeq\ker C_6(\zeta)$, where $C_6(\zeta)$ is
the $144\times144$ compatibility matrix defined by \eqref{eq:6local} and
\eqref{eq:6compat}.
\end{proposition}
\begin{proof}
At each target $j$, Lemma~\ref{lem:6quadrics} uniquely parametrizes
$(E_6\ell)_j=0$ by
\begin{equation}\label{eq:6local}
g_i^{(j)}:=w_{ji}\overline\ell_i
=\sum_{h\in S_j\setminus\{i\}}\overline Q_hc_{ih}^{(j)},
\qquad c_{hi}^{(j)}=-c_{ih}^{(j)},
\end{equation}
using six linear forms in $Y/W_j$ ($24$ parameters per target, $144$ in
total). For each input $i$, its four targets are
$J_i=\{0,\ldots,5\}\setminus\{i,i-1\}$. Since every $w_{ji}\ne0$ in
$F(\zeta)$, Lemma~\ref{lem:6gluing} applied to $w_{ji}^{-1}g_i^{(j)}$
shows that these local cubics glue to a unique $\ell_i\in\Sym^3Y$ if and
only if
\begin{equation}\label{eq:6compat}
\begin{aligned}
w_{ki}g_i^{(j)}-w_{ji}g_i^{(k)}&=0\quad\text{modulo }W_j+W_k,\\
j,k&\in J_i,\quad j<k.
\end{aligned}
\end{equation}
Six inputs, six target pairs per input, and four coordinates per binary
cubic give $144$ equations, yielding a $144\times144$ matrix $C_6(\zeta)$
with
\begin{equation}\label{eq:6keriso}
\ker E_6\simeq\ker C_6(\zeta).
\end{equation}
\end{proof}

\begin{lemma}[Normalized leading defect]\label{lem:6leading}
Normalize the compatibility rows as in \eqref{eq:6normalized} and write
$\widetilde C_6=C_0+\zeta C_1+\zeta^2C_2+\zeta^3C_3$. Then $C_0$ is a
direct sum of six $24\times24$ target blocks, each with the
two-dimensional kernel \eqref{eq:6smallkernel}; hence $\rank C_0=132$
and $\dim\ker C_0=\dim\coker C_0=12$.
\end{lemma}
\begin{proof}
For input $i$, the target $\zeta$-valuations $\ord_\zeta(w_{ji})$ are
ordered as
\begin{equation}\label{eq:6order}
i-3\ (0),\quad i-2\ (1),\quad i-4\ (2),\quad i-5\ (3).
\end{equation}
Dividing each equation \eqref{eq:6compat} by the smaller $\zeta$-power
and orienting it as
\begin{equation}\label{eq:6normalized}
g_i^{(\mathrm{high})}
-\zeta^{v_{\mathrm{high}}-v_{\mathrm{low}}}g_i^{(\mathrm{low})}=0
\end{equation}
yields $\widetilde C_6=C_0+\zeta C_1+\zeta^2C_2+\zeta^3C_3$, where $C_0$
splits into six target blocks. Fix target $j$, abbreviate
\[
\begin{aligned}
Z&=j+1,& A&=j+2,& B&=j+3,\\
C&=j+4,& D&=j+5,
\end{aligned}
\]
and work in $V=Y/W_j$. Any two projected planes $W_h$ are complementary,
and in $V/W_h$ any two remaining $Q_i$ are coprime binary quadrics.
Reading \eqref{eq:6order} gives the $24$ leading conditions at $j$:
\begin{equation}\label{eq:6leadingconditions}
\begin{array}{c|c|r}
\text{input}&\text{zero quotients}&\text{coords.}\\\hline
g_A&V/W_D&4\\
g_B&\text{none}&0\\
g_C&V/W_Z,\ V/W_A&8\\
g_D&V/W_A,\ V/W_B,\ V/W_Z&12
\end{array}
\end{equation}
In a binary ring, $Pu+Qv=0$ with coprime quadrics $P,Q$ and linear $u,v$
forces $u=v=0$ since $P\mid v$ and $\deg v<\deg P$.

Applying this observation to $g_D=Q_Ac_{DA}+Q_Bc_{DB}+Q_Cc_{DC}$ modulo
$W_A,W_B,W_Z$ in turn forces $c_{DC}\in W_A\cap W_B=0$,
$c_{DA}\in W_B\cap W_Z=0$, and $c_{DB}\in W_A\cap W_Z=0$. Thus the
twelve coordinates of $c_{D*}$ map isomorphically to the twelve $g_D$
outputs. With $c_{D*}=0$, $g_C=Q_Ac_{CA}+Q_Bc_{CB}$ modulo $W_A$ and
$W_Z$ gives $c_{CB}\in W_A\cap W_Z=0$ and $c_{CA}\in W_Z$. Finally
$g_A=Q_Bc_{AB}-Q_Cc_{CA}\equiv0\pmod{W_D}$ forces
$c_{CA}\in W_Z\cap W_D=0$ and $c_{AB}\in W_D$. Hence the exact block
kernel is
\begin{equation}\label{eq:6smallkernel}
\begin{aligned}
c_{AB}&=z_j\in W_D,\qquad g_C=g_D=0,\\
g_A&=Q_Bz_j,\qquad g_B=-Q_Az_j,
\end{aligned}
\end{equation}
with all other parameters zero. Thus each block has kernel dimension $2$
and rank $22$, so $\rank C_0=132$ and $\dim\ker C_0=\dim\coker C_0=12$.
\end{proof}

\begin{lemma}[Cokernel coordinates and six first-order isomorphisms]\label{lem:6firstorder}
After eliminating the $g_D$ outputs of each leading block, its cokernel
is the two-dimensional quotient $H_j$ in \eqref{eq:6smallcokernel}.
In these coordinates, $C_1$ induces $T_j$ in \eqref{eq:6T} from source
block $j+2$ to target block $j$. Every $T_j$ is an isomorphism, so the
induced $12$-dimensional map $\ker C_0\to\coker C_0$ is invertible in
every characteristic.
\end{lemma}
\begin{proof}
Since $c_{D*}$ maps isomorphically to the twelve $g_D$ outputs,
eliminating $g_D$ leaves a $12\times12$ reduced map on
$c_{AB},c_{CA},c_{CB}$. Put $U_j=Y/(W_j+W_A)$. The $g_C$ output modulo
$W_A$ of the reduced map is $Q_Bc_{CB}\bmod W_A$, so its class in
\begin{equation}\label{eq:6smallcokernel}
H_j=\Sym^3U_j/(Q_B\Sym^1U_j)
\end{equation}
annihilates the image. Because $\dim H_j=4-2=2$ equals the reduced
cokernel dimension and the projection onto $H_j$ is surjective, this
projection identifies the entire cokernel.

In \eqref{eq:6normalized}, a row has a $\zeta^1$ term precisely when
$v_{\mathrm{high}}-v_{\mathrm{low}}=1$, in which case $C_1=[\zeta^1]\widetilde C_6$
places $-g_i^{(\mathrm{low})}\bmod(W_{j_{\mathrm{high}}}+W_{j_{\mathrm{low}}})$
into the output slot of $g_i^{(\mathrm{high})}\bmod W_{j_{\mathrm{low}}}$
at target block $j=j_{\mathrm{high}}$. By \eqref{eq:6order}, each block
$j$ is $j_{\mathrm{high}}$ for exactly three consecutive valuation pairs:
\begin{itemize}
\item $(v_{\mathrm{low}},v_{\mathrm{high}})=(0,1)$: here $j=i-2$, so
$i=j+2=A$ and $j_{\mathrm{low}}=i-3=j-1$. At source block $j-1$, input
$i=(j-1)+3$ is the local $B$-slot, where \eqref{eq:6smallkernel} gives
$g_B^{(j-1)}=-Q_{(j-1)+2}z_{j-1}=-Q_Zz_{j-1}$, contributing
$-g_i^{(j-1)}=+Q_Zz_{j-1}$ to the slot $g_A\bmod W_{j-1}=g_A\bmod W_D$
(since $j-1\equiv j+5=D\pmod6$).
\item $(v_{\mathrm{low}},v_{\mathrm{high}})=(1,2)$: here $j=i-4$, so
$i=j+4=C$ and $j_{\mathrm{low}}=i-2=j+2=A$. At source block $j+2$, input
$i=(j+2)+2$ is the local $A$-slot, where \eqref{eq:6smallkernel} gives
$g_A^{(j+2)}=Q_{(j+2)+3}z_{j+2}=Q_Dz_{j+2}$, contributing
$-g_i^{(j+2)}=-Q_Dz_{j+2}$ to the slot $g_C\bmod W_A$.
\item $(v_{\mathrm{low}},v_{\mathrm{high}})=(2,3)$: here $j=i-5$, so
$i=j+5=D$ and $j_{\mathrm{low}}=i-4=j+1$. At source block $j+1$, input
$i=(j+1)+4$ is the local $C$-slot, where $g_C^{(j+1)}=0$ on
\eqref{eq:6smallkernel}, so all $g_D$ first-order entries vanish.
\end{itemize}
Collecting the nonzero entries at target $j$ gives
\begin{equation}\label{eq:6firstoutputs}
\begin{array}{c|c|c}
\text{source}&\text{output position}&\text{value}\\\hline
z_{j-1}&g_A\bmod W_D&+Q_Zz_{j-1}\\
z_{j+2}&g_C\bmod W_A&-Q_Dz_{j+2}
\end{array}
\end{equation}
Because all $g_D$ entries of $C_1$ on $\ker C_0$ are zero, no
$g_D$-elimination correction is triggered, and projecting onto $H_j$
drops the $g_A\bmod W_D$ row of \eqref{eq:6firstoutputs} while keeping
the $g_C\bmod W_A$ row:
\begin{equation}\label{eq:6T}
T_j:z_{j+2}\longmapsto[-Q_Dz_{j+2}]\in H_j.
\end{equation}
At its source, $z_{j+2}$ lies in $W_{(j+2)+5}\bmod W_{j+2}=W_Z\bmod W_A$
and lifts uniquely to $W_Z$, which maps isomorphically to $U_j$ by
six-node independence. In $\Sym U_j$, the quadrics $Q_B,Q_D$ are coprime,
so $Q_Dz=Q_Bu$ with linear $z,u$ forces $Q_B\mid z$ and hence $z=0$.
Thus each $T_j$ is an isomorphism between $2$-dimensional spaces, and
the global induced map is invertible as a block permutation of these six
isomorphisms.
\end{proof}

\subsection{Two Schur lifts complete the determinant proof}
\begin{proof}[Proof of Theorem~\ref{thm:higherrepair} for $t=6$]
By Lemmas~\ref{lem:6leading}--\ref{lem:6firstorder} and
Lemma~\ref{lem:lift}, $\det\widetilde C_6$ has a nonzero $\zeta^{12}$
leading term. For each input, the sum of smaller valuations over its six
target pairs is $0+0+0+1+1+2=4$, so undoing the row normalizations
multiplies the determinant by $\pm\zeta^{6\cdot4\cdot4}=\pm\zeta^{96}$,
giving a nonzero $\zeta^{108}$ term in $\det C_6(\zeta)$.
Proposition~\ref{prop:6equivalence} then implies that the $336\times336$
map $E_6$ is invertible over $F(\zeta)$. Applying Lemma~\ref{lem:lift}
a second time via Proposition~\ref{prop:6paired} produces a nonzero
$\eta^{336}$ term in the specialized repair determinant over
$F(\zeta,\eta)$, proving polynomial nonvanishing over $F$ in every
characteristic.
\end{proof}

\bibliographystyle{IEEEtran}
\bibliography{MSR_matrix_explicit_references}

\begin{thebibliography}{10}
\providecommand{\url}[1]{#1}
\csname url@samestyle\endcsname
\providecommand{\newblock}{\relax}
\providecommand{\bibinfo}[2]{#2}
\providecommand{\BIBentrySTDinterwordspacing}{\spaceskip=0pt\relax}
\providecommand{\BIBentryALTinterwordstretchfactor}{4}
\providecommand{\BIBentryALTinterwordspacing}{\spaceskip=\fontdimen2\font plus
\BIBentryALTinterwordstretchfactor\fontdimen3\font minus
  \fontdimen4\font\relax}
\providecommand{\BIBforeignlanguage}[2]{{%
\expandafter\ifx\csname l@#1\endcsname\relax
\typeout{** WARNING: IEEEtran.bst: No hyphenation pattern has been}%
\typeout{** loaded for the language `#1'. Using the pattern for}%
\typeout{** the default language instead.}%
\else
\language=\csname l@#1\endcsname
\fi
#2}}
\providecommand{\BIBdecl}{\relax}
\BIBdecl

\bibitem{survey}
V.~Ramkumar, S.~B. Balaji, B.~Sasidharan, M.~Vajha, M.~N. Krishnan, and P.~V.
  Kumar, ``Codes for distributed storage,'' \emph{Foundations and Trends in
  Communications and Information Theory}, vol.~19, no.~4, pp. 547--813, 2022.

\bibitem{dimakis}
A.~G. Dimakis, P.~B. Godfrey, Y.~Wu, M.~J. Wainwright, and K.~Ramchandran,
  ``Network coding for distributed storage systems,'' \emph{IEEE Transactions
  on Information Theory}, vol.~56, no.~9, pp. 4539--4551, 2010.

\bibitem{pm}
K.~V. Rashmi, N.~B. Shah, and P.~V. Kumar, ``Optimal exact-regenerating codes
  for distributed storage at the {MSR} and {MBR} points via a product-matrix
  construction,'' \emph{IEEE Transactions on Information Theory}, vol.~57,
  no.~8, pp. 5227--5239, 2011.

\bibitem{ybaccess}
M.~Ye and A.~Barg, ``Explicit constructions of optimal-access {MDS} codes with
  nearly optimal sub-packetization,'' \emph{IEEE Transactions on Information
  Theory}, vol.~63, no.~10, pp. 6307--6317, 2017.

\bibitem{wtb}
Z.~Wang, I.~Tamo, and J.~Bruck, ``Explicit minimum storage regenerating
  codes,'' \emph{IEEE Transactions on Information Theory}, vol.~62, no.~8, pp.
  4466--4480, 2016.

\bibitem{wlhy}
N.~Wang, G.~Li, S.~Hu, and M.~Ye, ``Constructing {MSR} codes with
  subpacketization $2^{n/3}$ for $k+1$ helper nodes,'' \emph{IEEE Transactions
  on Information Theory}, vol.~69, no.~6, pp. 3775--3792, 2023.

\bibitem{lwhy}
G.~Li, N.~Wang, S.~Hu, and M.~Ye, ``{MSR} codes with linear field size and
  smallest sub-packetization for any number of helper nodes,'' \emph{IEEE
  Transactions on Information Theory}, vol.~70, no.~11, pp. 7790--7806, 2024.

\bibitem{lts2026}
\BIBentryALTinterwordspacing
J.~Li, X.~Tang, and L.~Song, ``New minimum storage regenerating codes with
  small sub-packetization levels for all admissible repair degrees,''
  arXiv:2610.15329v1, Oct. 11, 2026. [Online]. Available:
  \url{https://arxiv.org/abs/2610.15329v1}
\BIBentrySTDinterwordspacing

\bibitem{smalld}
M.~Vajha, S.~B. Balaji, and P.~V. Kumar, ``Small-$d$ {MSR} codes with optimal
  access, optimal sub-packetization, and linear field size,'' \emph{IEEE
  Transactions on Information Theory}, vol.~69, no.~7, pp. 4303--4332, 2023.

\bibitem{bvk}
S.~B. Balaji, M.~Vajha, and P.~V. Kumar, ``Lower bounds on the
  sub-packetization level of {MSR} codes and characterizing optimal-access
  {MSR} codes achieving the bound,'' \emph{IEEE Transactions on Information
  Theory}, vol.~68, no.~10, pp. 6452--6471, 2022.

\bibitem{ag}
O.~Alrabiah and V.~Guruswami, ``An exponential lower bound on the
  sub-packetization of minimum storage regenerating codes,'' \emph{IEEE
  Transactions on Information Theory}, vol.~67, no.~12, pp. 8086--8093, 2021.

\bibitem{elyasimohajer2019}
M.~Elyasi and S.~Mohajer, ``Determinant codes with helper-independent repair
  for single and multiple failures,'' \emph{IEEE Transactions on Information
  Theory}, vol.~65, no.~9, pp. 5469--5483, 2019.

\bibitem{elyasimohajer2020}
------, ``Cascade codes for distributed storage systems,'' \emph{IEEE
  Transactions on Information Theory}, vol.~66, no.~12, pp. 7490--7527, 2020.

\bibitem{duursmaliwang2020}
I.~M. Duursma, X.~Li, and H.-P. Wang, ``Multilinear algebra for distributed
  storage,'' \emph{SIAM Journal on Applied Algebra and Geometry}, vol.~5,
  no.~3, pp. 552--587, 2021.

\bibitem{eidduursma2016}
\BIBentryALTinterwordspacing
A.~Eid and I.~Duursma, ``A function field approach to construct {MBR} and {MSR}
  regenerating codes with auxiliary data,'' Author manuscript, Dec. 21, 2016.
  [Online]. Available:
  \url{https://www.abdullaeid.net/teaching/msrmbrpolynomialFeb29.pdf}
\BIBentrySTDinterwordspacing

\bibitem{atrahasis}
I.~Duursma and H.-P. Wang, ``Multilinear algebra for minimum storage
  regenerating codes: a generalization of the product-matrix construction,''
  \emph{Applicable Algebra in Engineering, Communication and Computing},
  vol.~34, no.~4, pp. 717--743, 2023, first published online in 2021.

\bibitem{patrabarg2025}
A.~Patra and A.~Barg, ``Generalized regenerating codes and node repair on
  graphs,'' \emph{IEEE Transactions on Information Theory}, vol.~71, no.~3, pp.
  1613--1630, 2025.

\bibitem{teamPatent2025}
\BIBentryALTinterwordspacing
K.~Wang, F.~Gao, R.~Li, Y.~Zhao, Y.~Zhang, X.~Li, X.~Chen, and G.~Xu, ``Data
  repair method, distributed storage system, device, medium and product,''
  Chinese patent application {CN~120215840~A}, Jun.~27, 2025, in Chinese.
  Applicant: Inspur Electronic Information Industry Co., Ltd. [Online].
  Available: \url{https://patents.google.com/patent/CN120215840A/zh}
\BIBentrySTDinterwordspacing

\bibitem{krickszanto2026}
T.~Krick and A.~Szanto, ``A note on the multivariate symmetric {Hermite}
  interpolant,'' \emph{Journal of Symbolic Computation}, vol. 134, p. 102505,
  2026.

\bibitem{vzgg}
J.~von~zur Gathen and J.~Gerhard, \emph{Modern Computer Algebra}, 3rd~ed.\hskip
  1em plus 0.5em minus 0.4em\relax Cambridge University Press, 2013.

\bibitem{glynn1986}
D.~G. Glynn, ``The non-classical 10-arc of {PG}(4,9),'' \emph{Discrete
  Mathematics}, vol.~59, no. 1--2, pp. 43--51, 1986.

\bibitem{hirschfeld1998}
J.~W.~P. Hirschfeld, \emph{Projective Geometries over Finite Fields},
  2nd~ed.\hskip 1em plus 0.5em minus 0.4em\relax Oxford, U.K.: Oxford
  University Press, 1998.

\end{thebibliography}
\end{document}